\documentclass[aps,prl,twocolumn,superscriptaddress,floatfix,nofootinbib]{revtex4-2}
\usepackage[english]{babel} 
\usepackage{amssymb}
\usepackage{amsmath}
\usepackage{mathtools}
\usepackage{txfonts}
\usepackage{mathdots}
\usepackage[normalem]{ulem}
\usepackage[dvips]{graphicx}
\usepackage{epsfig}
\usepackage{graphicx}
\usepackage{array}
\usepackage{amssymb}
\usepackage{amsfonts}
\usepackage{amsmath}
\usepackage{mathrsfs}
\usepackage{booktabs}
\usepackage{threeparttable}
\usepackage{multirow}
\usepackage{subfigure}
\usepackage{epsfig}
\usepackage{threeparttable}
\usepackage{chngpage}
\usepackage{float}
\usepackage{bm}
\usepackage{graphicx}
\usepackage{adjustbox}
\usepackage{comment}
\usepackage{hyperref}
\usepackage{tabularx}
\usepackage{graphicx}
\usepackage{adjustbox}
\usepackage[dvipsnames]{xcolor}

\hypersetup{
    unicode=false,     
    pdftoolbar=false,  
    pdfmenubar=true,   
    pdffitwindow=false, 
    pdfstartview={FitH},
    pdftitle={},    
    pdfauthor={Authors},     
    pdfsubject={},   
    pdfcreator={},   
    pdfproducer={}, 
    pdfkeywords={} {} {}, 
    pdfnewwindow=true,
    colorlinks=true,
    linkcolor=black,
    citecolor=blue, 
    filecolor=magenta,
    urlcolor=blue
}

\newcommand{\beginsupplement}{%
    \setcounter{table}{0}
    \renewcommand{\thetable}{S\arabic{table}}%
    \setcounter{figure}{0}
    \renewcommand{\thefigure}{S\arabic{figure}}%
    \setcounter{equation}{0}
    \renewcommand{\theequation}{S\arabic{equation}}%
    \setcounter{section}{0}
    \renewcommand{\thesection}{S\arabic{section}}%
   }

\makeatletter

\newcommand{\Rmnum}[1]{\expandafter\@slowromancap\romannumeral #1@}
\makeatother

\newcommand{\br}[1]{\langle#1\rangle}

\newcommand{\eps}{\epsilon}

\newcommand{\Zb}{\mathbb{Z}}

\newcommand{\sbr}[1]{\left[#1\right]}

\newcommand{\pd}[2]{\ensuremath{\frac{\partial#1}{\partial#2}}}

\newcommand{\md}[6]{\ensuremath{
		\ifinner
		\tfrac{\partial{^{#2}}#1}{\partial{#3^{#4}}\partial{#5^{#6}}}
		\else
		\tfrac{\partial{^{#2}}#1}{\partial{#3^{#4}}\partial{#5^{#6}}}
		\fi
}}

\newcommand{\1}{\mathds{1}}

\newcommand{\BH}{{\mathrm{BH}}}

\begin{document}

\title{
Disorder-Induced Logarithmic Spreading and Localization of Quantum Fluctuations
}

\author{Ang Yang}
\affiliation{School of Physics, and Zhejiang Province Key Laboratory of Quantum Technology and Device, Zhejiang University, Hangzhou $310027$, China}

\author{Jingxuan Zhang}
\email{jingxuan@tsinghua.edu.cn}
\affiliation{Yau Mathematical Sciences Center, Tsinghua University, Beijing $100084$}
\affiliation{Beijing Institute of Mathematical Sciences and Applications, Beijing $101408$}

\author{Lei Ying}
\email{leiying@zju.edu.cn}
\affiliation{School of Physics, and Zhejiang Province Key Laboratory of Quantum Technology and Device, Zhejiang University, Hangzhou $310027$, China}

\begin{abstract}
Disorder can localize quantum particles, even in the presence of interactions, but its influence on quantum fluctuations beyond mean-field dynamics remains poorly understood. Here, we investigate the propagation of quantum fluctuations in weakly interacting disordered bosonic systems by comparing the Bose–Hubbard model with the Gross–Pitaevskii equation. We find that, in contrast to ballistic spreading in the disorder-free system, quantum fluctuations exhibit logarithmic spreading at intermediate times under both random and quasiperiodic disorder, before eventually saturating. The spreading rate decreases algebraically with increasing disorder strength, and this behavior persists across different particle numbers and extends to two dimensions. Using a time-dependent Bogoliubov–de Gennes framework, we identify the localized mean-field wavefunction as a time-dependent pairing background that generates quantum fluctuations. The exponential suppression of fluctuation-generation rates with distance provides a physical mechanism for the logarithmically advancing fluctuation front. Our results reveal a distinct regime of disorder-controlled quantum-fluctuation dynamics and establish disorder as a means of controlling their spreading.
\end{abstract}
\date{\today}
\maketitle

\emph{Introduction}---Quantum fluctuations are an intrinsic feature of quantum systems arising from the non-commutativity of quantum observables and the uncertainty principle. They play essential roles in quantum phase transitions~\cite{Sachdev_2011,Fisher1989prb,Greiner2002SF}, quantum criticality~\cite{Hertz1976critical,coleman2005criticality}, and nonequilibrium dynamics~\cite{berges2007pra,Trefzger2011prl,Lovas2017pra}. In interacting many-body systems, quantum fluctuations characterize deviations from mean-field descriptions and can develop spatial correlations and complex dynamics beyond mean-field theory~\cite{castin1997instability,castin1998bdg,thierry2003onedimension,Pitaevskii2016,Silke2021}. On the theoretical side, the rigorous derivation of mean-field theories and the characterization of quantum corrections have been extensively investigated, following foundational works by Spohn~\cite{Spohn1980}, Erdos–Schlein–Yau~\cite{erdHos2009rigorous,erdos2010derivation}, and numerous subsequent developments~\cite{AKS,BPS,benedikter2015quantitative,brennecke2019gross,bossmann2022beyond,chen2016klainerman,chen2019derivation,deuchert2023dynamics,DL,GMM,GMMa,knowles2010mean,Kuz,RL23,lewin2015fluctuations,mitrouskas2019bogoliubov,nam2017bogoliubov,pickl2011simple,RS,RS22,napiorkowski2023dynamics}. Understanding how quantum fluctuations propagate in many-body systems therefore remains a fundamental question.

Disorder provides a powerful mechanism for controlling quantum transport and localization. In noninteracting systems, quantum interference suppresses particle transport and leads to Anderson localization~\cite{anderson1958,anderson1979scaling}. In interacting systems, the interplay between disorder and interactions can give rise to many-body localization, characterized by the absence of thermalization, persistent memory of initial conditions, and slow growth of entanglement and correlations~\cite{huse2007prb,huse2010prb,huse2015annual,john2016prl,bloch2015science,papic2015MBL,abanin2019reviewMBL}. Although disorder-induced localization has been extensively investigated~\cite{anderson1958,Bilas2006Bogoliubov,Aspect2007AL,Lugan2007Bogoliubov,Pavloff2007superfluidity,abanin2019reviewMBL,yang2026origin}, how disorder affects the propagation of quantum fluctuations beyond mean-field dynamics remains poorly understood.

In this work, we investigate the propagation of quantum fluctuations beyond mean-field dynamics in weakly interacting disordered bosonic systems by comparing the Bose–Hubbard (BH) model with the Gross–Pitaevskii (GP) equation, as illustrated in Fig.~\ref{schematic}(a). We identify an intermediate-time regime in which quantum fluctuations spread logarithmically, even though the mean-field wavefunction remains spatially localized, before eventually saturating [Fig.~\ref{schematic}(b)]. This behavior occurs under both random and quasiperiodic disorder, with the spreading rate decreasing as the disorder strength increases. It persists across different particle numbers and extends to two dimensions. To elucidate the underlying mechanism, we develop a time-dependent Bogoliubov–de Gennes (BdG) framework. We show that the localized GP wavefunction acts as a time-dependent pairing background that generates quantum fluctuations. The resulting exponential suppression of fluctuation-generation rates with distance provides a physical explanation for the logarithmically advancing fluctuation front.

\begin{figure}
\centering
\includegraphics[width=\linewidth]{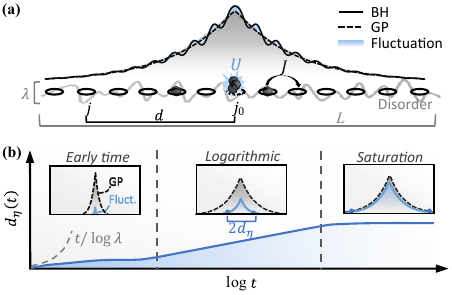}
\caption{
(a) Schematic of the many-body and mean-field density distribution, and the quantum fluctuation in a disordered one-dimensional Bose-Hubbard chain. All the bosons are initially condensed at the central site $j_0=0$, while the local operator probes a site $j$ at distance $d=|j-j_0|$. (b) Schematic dynamics of the fluctuation front $d_\eta(t)$, defined as the distance from $j_0$ to a fixed amplitude of the disorder-averaged fluctuation density distribution $\eta$. The front exhibits an early-time ballistic bound, intermediate-time logarithmic spreading, and eventual saturation.
Insets in three time regimes illustrate the distributions of the GP density and the fluctuation density. Blue dots mark a fixed fluctuation density value $\eta$.
}
\label{schematic}
\end{figure}

\emph{Models and fluctuations}---We consider weakly interacting bosons in the lattice $\Zb^\nu$ with dimension $\nu\ge1$, subjected to a disordered external potential, described by the BH Hamiltonian 
\begin{equation}\label{eq:Hintro}
\hat{H}_\BH=-J \sum_{\br{i,j}} \hat{b}_{i}^{\dagger} \hat{b}_{j}+\frac{U}{2 N} \sum_{i\in R} \hat{n}_i\left(\hat{n}_i-1\right)+\sum_{i\in R} \lambda_i \hat{b}_i^{\dagger} \hat{b}_i.
\end{equation}
Here, ${\br{i,j}}$ denotes pairs of neighboring sites, $J$ is the hopping amplitude, $N$ is the number of bosons, $U$ is the interaction strength, and $\lambda_i$ is the disordered potential strength at site $i$. We consider periodic boundary condition and $R$ denotes the $\nu$- dimensional box of side length $L$ around the origin.

The crux of the mean-field approximation is that, starting with a purely factorized (i.e., condensate) initial state,  for   $N$ sufficiently large, the  Schr\"odinger equation associated with \eqref{eq:Hintro} is well approximated by the GP equation~\cite{Gross1961,Pitaevskii2016},
\begin{equation}\label{NLS}
\begin{split}
i\pd{\psi(t)}{t} &= \hat{H}_\mathrm{GP}(t) \psi(t), \\
 \hat{H}_{\mathrm{GP},ij}(t)&=-J (\delta_{i,j+1}+\delta_{i,j-1}) + (\lambda_i+U\lvert\psi_i(t)\rvert^2) \delta_{ij}.
 \end{split}
\end{equation}
In passing from \eqref{eq:Hintro} to the effective equation \eqref{NLS}, the degree of freedom reduces from $N \nu$ to just $\nu$, at the cost of an extra cubic nonlinearity.

\begin{figure*}
\centering
\includegraphics[width=\linewidth]{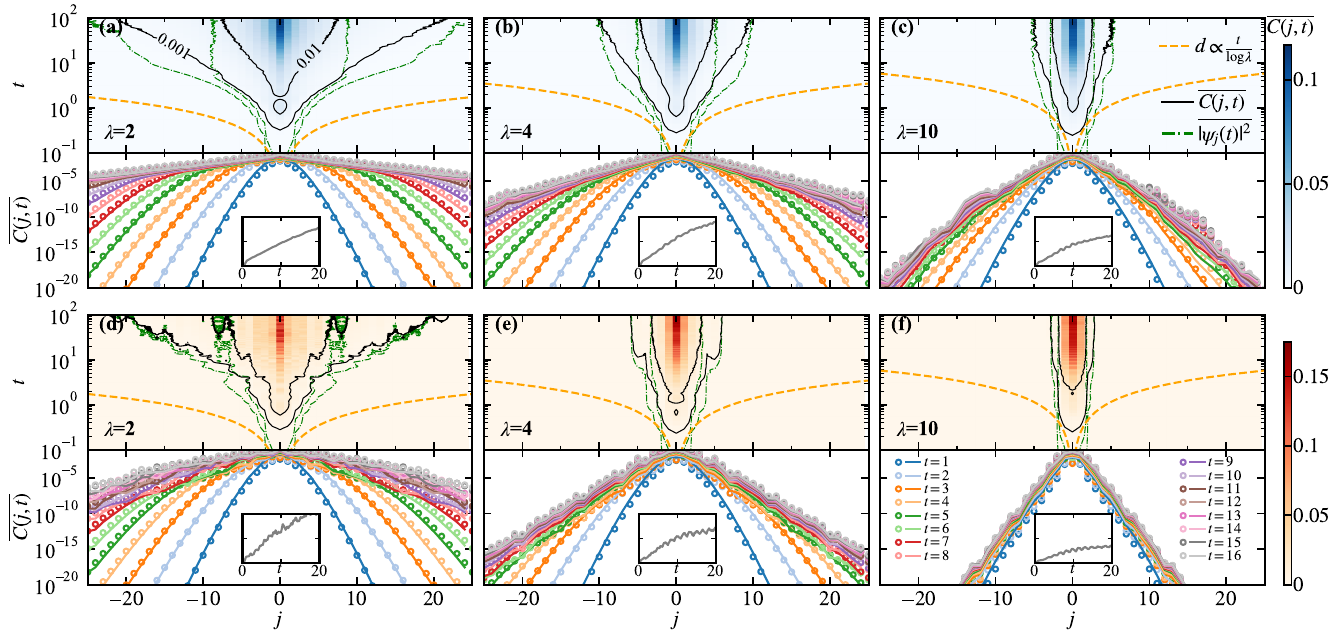}
\caption{
(a-c) The averaged quantum fluctuation density distribution $\overline{C(j,t)}$ as a function of lattice index $j$ and time $t$ for different random disorder strengths $\lambda$. (d-f) Similar data as in (a-c) but in a quasiperiodic disorder setting. In each panel, the lower half shows the corresponding spatial profiles of $\overline{C(j,t)}$ at different early times $t$ on a semilog scale, where the circles denote the corresponding BdG results. Insets show the dynamics of the averaged value of $\overline{C(j,t)}$ in $j\in[-10,10]$ on a linear scale. The black solid lines denote the contours at $\overline{C(j,t)}=10^{-3}$ and $10^{-2}$, while the green dash-dotted lines denote the same contour of the density distribution $\overline{|\psi_j(t)|^2}$ of the GP wavefunction. The orange-dashed lines denote the estimated ballistic bound of transport $d\propto t/\mathrm{log}\lambda$. The particle number and the system size are $N=4$ and $L=50$, respectively. The BdG results are averaged over $N_{\mathrm{s}}=1000$ realizations with a system size $L=100$.
}
\label{transport}
\end{figure*}

To quantify this approximation, let $\phi(t)$ be a Schr\"odinger orbit generated by the BH Hamiltonian \eqref{eq:Hintro}, and  $\psi(t)$ be a solution to the GP equation~\eqref{NLS}. Define the one-particle reduced density matrix $\rho(t)$ in the second quantization form as 
$\rho_{ij}(t)=(1/N)\langle \phi\lvert\hat{b}_i^{\dagger}(t) \hat{b}_j(t)\rvert\phi\rangle$.
The correction of mean-field approximation is quantified by  the \emph{quantum fluctuation}~\cite{RS} 
\begin{equation}
Q(t)=\rho (t)-   \vert \psi(t) \rangle \langle \psi(t) \vert. 
\end{equation}
To illustrate the local enhancement of the mean-field approximation, we consider the geometric setup as depicted in Fig.~\ref{schematic}(a). We probe the BH and the GP models with the \textit{local} observable $\hat O_j$ at site $j$. A typical and experimentally feasible local observable~\cite{Greiner2002SF,bloch2015science,choi2016exploring,Greiner2019science} is the local density operator, $\hat n_j=\hat b_j^\dagger\hat b_j$. Thus, we can measure local fluctuation density
\begin{equation}
   C(j,t)=\Big|\mathrm{Tr}\big(Q(t)\hat{n}_j\big)\Big|,
   \label{Cdef}
\end{equation} 
which demonstrates the difference in particle number at each site between the BH and GP model. 
At $t=0$, assume the condensate structure $\phi(0)=\sbr{\psi(0)}^{\otimes N}$ holds with the condensate wave function $\psi(0)=\delta_{j_0}$ at a given site $j_0$.

\begin{figure}
\centering
\includegraphics[width=1\linewidth]{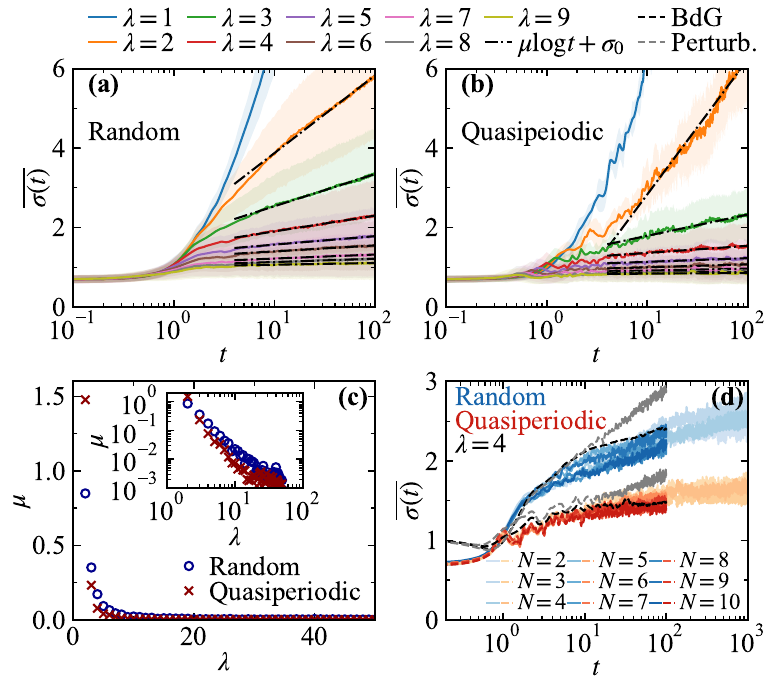}
\caption{
(a-b) The averaged RMSD $\overline{\sigma(t)}$ as a function of time $t$ on a semilog scale, for different random (a) and quasiperiodic (b) disorder strengths $\lambda$. The black dash-dotted lines denote the fitting curve $\mu\mathrm{log}t+\sigma_0$, and the color-shaded area denotes the standard deviation at each disorder strength. (c) The fitted slope $\mu$ as a function of disorder strength $\lambda$ for random and quasiperiodic settings, and the inset shows the same data on a log-log scale. The particle number and the system size are $N=4$ and $L=50$, respectively. (d) The averaged RMSD $\overline{\sigma(t)}$ at $\lambda=4$ as a function of time $t$ for different particle numbers $N$ and disorder settings. For results at $N=2\text{-}10$, the sampling number and system size are $N_{\mathrm{s}}=100$ and $L=150, 100, 50, 50, 100, 100$. We use TEBD for $N=6\text{-}10$ with the maximum bond dimension $\chi=1024$, and time step $\delta t=0.01$ with a second-order Trotter decomposition. The black-dashed lines denote the BdG results, sampled over $N_{\mathrm{s}}=1000$ realizations with a system size $L=100$, and the gray-dashed lines denote the perturbative results with $L=100, N_\mathrm{s}=100$.
}
\label{std}
\end{figure}

\emph{Slow spreading of fluctuation}---In the following, we first consider the one-dimensional (1D) case to investigate the transport of quantum fluctuation in both quasiperiodic and random settings. 
For the random setting, $\lambda_i$ is sampled from a uniform distribution in $[-\lambda,\lambda]$;
for the quasiperiodic setting, we consider $\lambda_i=\lambda \cos \big\{2\pi[(\sqrt{5}-1)i/2 +\theta]\big\}$ with $\theta$ sampled from a uniform distribution in $[0,1]$ ~\cite{wang2021critical}. Unless stated otherwise, we typically average the results over $N_{\mathrm{s}}=1000$ disorder realizations to ensure the convergence of quantum fluctuation~\cite{papic2015MBL}, and $\overline{\cdots}$ denotes the averaged quantities. We have checked that $N_{\mathrm{s}}\geq100$ is reliable to reach convergence~\cite{SM}. We set $J=U=1$ and consider the low-filling case ($N\ll L$). Our numerical results are obtained using either the exact evolution of the BH and GP models, or the time-evolving block decimation (TEBD) algorithm implemented in the open-source library TeNPy~\cite{Johannes2018tenpy}.

In Fig.~\ref{transport}, we show the spatiotemporal distributions of the fluctuation density $\overline{C(j,t)}$ for different disorder strengths $\lambda$ in the random (a-c) and quasiperiodic settings (d-f), respectively. At $t=0$, the fluctuation is absent due to the condensate structure, and therefore $C(j,0)\equiv 0$. At early times, the fluctuation emerges around the condensate $j_0$, a hallmark of approximation error between the many-body and mean-field dynamics. For better visibility of transport, we mark the contours at $\overline{C(j,t)}=10^{-2}, 10^{-3}$ as black lines. At short times $t\sim 1$, the propagation of quantum fluctuation gradually slows down as the disorder strength increases. This suppressed spreading behavior of quantum fluctuation is much more pronounced for the contour at $\overline{C(j,t)}=10^{-2}$ in the quasiperiodic case [Fig.~\ref{transport}(d-f)]. This behavior is further confirmed by the spatial profile, where $\overline{C(j,t)}$ remains strongly confined during the time evolution, and even exhibits an exponential decay in the strong-disorder regime [Fig.~\ref{transport}(f)]. We note that, in all cases, the fluctuation transport is well bounded by the recent theoretical estimate
$C(j,t) \leq (M/N) \big[ e^{K(t-{d}/{\eps})}+{e^{Kt}}/{N}\big]$  for $t\leq d/\epsilon$ \cite{LRZa,LRZb}. Here, $K,\,M,\,\eps$ are all independent of $N$, and $\eps$ can be made arbitrarily small for sufficiently strong disorder. Specifically, for a given disorder strength $\lambda$, one can choose $\epsilon$ to be of the order $1/\mathrm{log}\lambda$ (orange-dashed lines).

At longer times, more importantly, a pronounced spreading cone emerges in the space-time plane. Its contour indicates a nearly logarithmic spreading $\propto\mathrm{log}\ t$, in contrast to the ballistic spreading observed in the disorder-free case~\cite{SM}. As the disorder strength increases, the spreading of fluctuation becomes slower and more restricted, but still demonstrates a similar logarithmic spreading behavior. Notably, the logarithmic spreading of quantum fluctuation is clearly observed when the mean-field wavefunction remains spatially localized (green dash-dotted contours), although the spatial extent of the fluctuation density distribution appears to be bounded by that of the mean-field density profile. Eventually, the spatial profile of $\overline{C(j,t)}$ gradually freezes over time and exhibits an approximately exponential decay~\cite{SM}. These phenomena are present in both random and quasiperiodic settings, \emph{albeit} that the quasiperiodic disorder exhibits a significantly stronger suppression effect on the fluctuation spreading compared to random disorder. This is seen by comparing Fig.~\ref{transport}(b-c) with (e-f), where the spreading cone is much narrower in the quasiperiodic setting.

\emph{Interpretation in BdG framework---}Quantum fluctuations about the condensate can be described within the BdG framework, which captures the condensate's quasiparticle excitations in the large-$N$ limit. To formulate the fluctuations in a number-conserving manner, we introduce the fluctuation operators $\hat{\Lambda}_i^\dagger$ and $\hat{\Lambda}_i$ acting on the bosonic Fock space, and retaining terms up to second order in the fluctuation operators yields a quadratic Hamiltonian. By noting $\hat{\bm{\Lambda}}^{(\dagger)} = [\dots,\hat{\Lambda}_j^{(\dagger)},\dots]^T$, the corresponding Heisenberg equations of motion give the time-dependent BdG equation (see \cite{castin1998bdg,SM} for definitions):
\begin{equation}
\begin{aligned}
i \frac{d}{dt} 
\begin{pmatrix}
    \hat{\bm{\Lambda}} \\
    \hat{\bm{\Lambda}}^{\dagger}
\end{pmatrix}
= \mathcal{H}(t)
\begin{pmatrix}
    \hat{\bm{\Lambda}} \\
    \hat{\bm{\Lambda}}^{\dagger}
\end{pmatrix}, \quad
\mathcal{H}(t)=\begin{pmatrix}
    \mathcal{A}(t) & \mathcal{B}(t) \\
    -\mathcal{B}(t)^* & -\mathcal{A}(t)^*
\end{pmatrix}, \label{eq:bdg}
\end{aligned}
\end{equation}
where
$\mathcal{A}(t)=\hat{H}_{\mathrm{GP}}(t)+U P(t)\mathrm{diag}\big[\lvert\psi_i(t)\rvert^2\big]P(t)$,
$\mathcal{B}(t)=UP(t)\mathrm{diag}\big[\psi^2_i(t)\big]P^*(t)$, and  $P(t) = I-\psi(t)\psi^\dagger(t)$. 
The anomalous pairing term $\mathcal{B}(t)$, which depends on the GP wavefunction through $U\psi_j^2(t)$, describes the generation of quantum fluctuations through pair creation. Importantly, the GP wavefunction acts as a time-dependent pairing medium, through which quantum fluctuations are generated and subsequently spread.
The local quantum fluctuation density distribution can be approximately written as~\cite{SM}
\begin{equation}
C(j,t)\approx \big\lvert\langle\hat{\Lambda}_j^{\dagger}(t)\hat{\Lambda}_j(t)\rangle\big\rvert/N.
\end{equation}
Using the same set of parameters, we obtain the BdG results (circles) presented in the lower panels of Fig.~\ref{transport}. The BdG results are in agreement with the numerical results and successfully capture the suppressed spatial profile.

The propagation of quantum fluctuation can be understood in a perturbative way. We assume that the GP wavefunction remains spatially localized, with an approximately stationary density profile but site-dependent evolving phases (see \cite{SW} for exact GP solutions satisfying these properties). This allows us to treat the normal BdG matrix $\mathcal A$ as time-independent while retaining the temporal dependence of the anomalous pairing interaction. We further consider that the normal eigenmodes $u_\alpha(j)$, satisfying $\mathcal A u_\alpha=\varepsilon_\alpha u_\alpha$, are exponentially localized, and treat the anomalous pairing interaction perturbatively. For an initial fluctuation vacuum, the local fluctuation occupation is then expressed as~\cite{SM}
$\Pi_j(t)=\langle\hat{\Lambda}_j^\dagger(t)\hat{\Lambda}_j(t) \rangle=\sum_\beta\big|\sum_\alpha u_\alpha(j)Y_{\alpha\beta}(t)\big|^2$,
where $Y_{\alpha\beta}(t)=-ie^{-i\varepsilon_\alpha t}\int_0^t ds\,g_{\alpha\beta}(s)e^{i(\varepsilon_\alpha+\varepsilon_\beta)s}$ denotes the perturbative pairing amplitude, with $g_{\alpha\beta}(t)=u^{\ast}_{\alpha}\mathcal{B}(t)u^{\ast}_{\beta}$. We numerically verify the accuracy of the above perturbative equation; see details in~\cite{SM}.

Under the adiabatic evolution of $P(t)$, assuming that disorder averaging suppresses interference between distinct pairing contributions and that the overlap-weighted detuning distribution is sufficiently smooth and nonzero near resonance, the occupation crosses over to a golden-rule regime, $\overline{\Pi_j(t)}\approx\Gamma_jt$ (evidenced by insets in Fig.~\ref{transport}), where $\Gamma_j$ denotes the effective local pairing-generation rate~\cite{SM}. The exponential localization of the GP wavefunction~\cite{Iomin2007Localization,Cheng2010MatterWave} and normal BdG eigenmodes motivates the estimate $\Gamma_j\approx\Gamma_0e^{-|j-j_0|/\ell_{\mathrm{eff}}}$, provided the resonant spectral weight preserves this spatial dependence. Here, $\ell_{\mathrm{eff}}$ is the effective decay length of the disorder-averaged pairing rate. By defining the excitation distance through a fixed fluctuation occupation $\eta$, we have the estimated fluctuation front 
\begin{equation}
    d_\eta(t)\approx \ell_\mathrm{eff}\log\left(\frac{\Gamma_0}{\eta}t\right). 
\end{equation}
This provides a physical picture for logarithmically advancing fluctuation contours arising from disorder-suppressed pairing processes.

\emph{RMSD dynamics---}To quantify the fluctuation spreading behavior and its localization, we consider the root mean-square displacement (RMSD), $\sigma(t)$, given by~\cite{Kloss2020,yang2024}
\begin{equation}
    \sigma(t) = \left[\sum\limits_{j=-L/2}^{L/2} \frac{j^2C(j,t)}{\mathcal{N}(t)} - \left(\sum\limits_{j=-L/2}^{L/2}\frac{jC(j,t)}{\mathcal{N}(t)} \right)^2\right]^{1/2},
\label{MSD}
\end{equation}
where the normalization factor is $\mathcal{N}(t)=\sum_jC(j,t)$, corresponding to the total fluctuations at time $t$. The RMSD $\sigma(t)$ captures the spreading distance $d$, and is expected to capture the behavior indicated by the contour in Fig.~\ref{transport}. We show the dynamics of the averaged RMSD for different disorder strengths in random (a) and quasiperiodic settings (b) in Fig.~\ref{std}. Larger disorder strength remarkably suppresses the growth of RMSD. It is noteworthy that RMSD indeed exhibits a logarithmic growth at intermediate times ($Jt\sim100$), as shown in the semilog-$t$ scale in Fig.~\ref{std}(a) and (b). Thus, we propose a logarithmic function $\overline{\sigma(t)}=\mu\mathrm{log}t+\sigma_0$ to quantitatively fit RMSD for different disorder strengths (black dash-dotted lines). This fitting function effectively captures the RMSD dynamics in either a random or quasiperiodic setting. The logarithmic growth of RMSD is also supported by the BdG analysis~\cite{SM}. We further investigate the suppression effect of disorder strength in terms of the fitted slope $\mu$. As shown in Fig.~\ref{std}(c), despite different disorder types, the fitted slope $\mu$ dramatically decreases towards zero as the disorder strength increases. 
Similar to the spreading behavior in Fig.~\ref{transport}, the quasiperiodic disorder exhibits a stronger suppression effect on the slope than the random case. Note that the inset in Fig.~\ref{std}(c) suggests that the slope drops algebraically with disorder strength at a moderate range. 
We notice that $\mu$ drops to the order of $10^{-3}$ around $\lambda=10$, indicating a potential localization of quantum fluctuation for large $\lambda$. The logarithmic growth of $\overline{\sigma(t)}$, together with the algebraic dependence of $\mu$ on $\lambda$, manifests a universal suppression effect of disorder on quantum fluctuation. 
A similar phenomenon is found starting from an initial condensate state with finite width~\cite{SM} $\psi(0)=(\delta_{j_0-1}+\delta_{j_0}+\delta_{j_0-1})/\sqrt{3}$.

\begin{figure}[t!]
\centering
\includegraphics[width=1\linewidth]{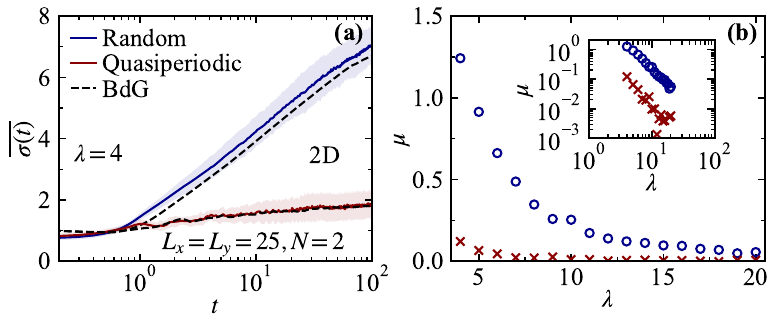}
\caption{
(a) The averaged RMSD $\overline{\sigma(t)}$ dynamics at $\lambda=4$ for the random (blue) and quasiperiodic (red) disorders in the 2D case. The black-dashed lines denote the BdG results, sampled over $N_{\mathrm{s}}=100$ realizations with a system size $L_x=L_y=30$. The color-shaded area denotes the standard deviation of RMSD. (b) The fitted slope $\mu$ as a function of disorder strength $\lambda$ for random and quasiperiodic settings. The particle number, the system size, and the sampling number are $N=2, L_x=L_y=25$, and $N_{\mathrm{s}}=100$.
}
\label{2d}
\end{figure}

Next, we demonstrate the RMSD dynamics at $\lambda=4$ for different particle numbers $N$ and different disorder types in Fig.~\ref{std}(d). At early times, the RMSD dynamics coincide closely for different $N$. After long-time evolution, the value of RMSD slightly decreases as $N$ increases. Importantly, the logarithmic growth rate is not altered by particle number. These behaviors are present in both disorder settings. Hence, these results suggest that disorder-induced suppression of quantum fluctuation holds uniformly for large $N$. 
Also, we show the BdG results (black-dashed lines), which reproduce the logarithmic growth at intermediate times.
Then, we perform a long-time evolution up to $Jt\sim10^3$ for the parameters considered here. At late times ($Jt>100$), particularly in the quasiperiodic case, the growth of RMSD slows down and saturates, indicating localization of quantum fluctuation in the long-time limit. For the random disorders, the RMSD gradually slows down its growth, and it still saturates for stronger disorders (see in~\cite{SM} for larger $\lambda$). This indicates that the logarithmic growth is a transient behavior, since the spatial extent of fluctuation $\overline{C(j,t)}$ is bounded by that of the density distribution $\overline{|\psi_j(t)|^2}$.

To further confirm the generality of the suppression effect of disorder in high dimensions, we show the RMSD dynamics at $\lambda=4$ for $N=2$ in a two-dimensional (2D) square lattice in Fig.~\ref{2d}(a). The logarithmic growth still exists up to $Jt\sim10^2$ in the two disorder settings (see in~\cite{SM} for other $\lambda$, where we also check the ballistic bound). 
Again, the BdG results (black-dashed lines) greatly reproduce the logarithmic growth at intermediate times.
Compared to the 1D case, the difference between the slope of RMSD dynamics in the two disorder settings is much more pronounced. After fitting the RMSD dynamics at different disorder strengths, we find a similar algebraic dependence of the slope $\mu$ on disorder strength as in the 1D case, as shown in Fig.~\ref{2d}(b). 
These results demonstrate that disorder-induced suppression of quantum-fluctuation spreading persists beyond one dimension.

\emph{Conclusions}---We have investigated disorder-induced suppression of quantum-fluctuation propagation in weakly interacting bosonic systems. For both random and quasiperiodic disorder, we verify a short-time ballistic propagation bound with arbitrarily small velocity and reveal robust logarithmic spreading at intermediate times. The logarithmic spreading rate decreases algebraically with disorder strength, with stronger suppression under quasiperiodic disorder. This behavior persists across different particle numbers and in two dimensions. A time-dependent BdG framework reproduces the observed dynamics and provides an intuitive understanding of logarithmic spreading through exponentially suppressed coupling to distant fluctuation modes. Our findings establish a distinct regime of disorder-controlled quantum-fluctuation dynamics, preceding eventual localization.

It is important to emphasize that the propagation bound we validate holds for all spatial dimensions and appears to be universal across various disorder regimes, as long as the probe operator is local and the initial condensate occupies finite sites. Interestingly, the logarithmic propagation and exponential decay structure of quantum fluctuation bear a resemblance to the growth of entanglement in many-body localization~\cite{abanin2019reviewMBL}, where interactions and disorder slow down the propagation of correlations. The analogy between these two phenomena raises intriguing questions about their connection, which warrants further exploration.

\emph{Acknowledgments}---We thank T.~Prosen, M.~Lemm, and S.~Rademacher for helpful discussions.  A.Y. and L.Y. were supported by the National Natural Science Foundation of China (No. 12375021), the Zhejiang Provincial Natural Science Foundation of China (No. LD25A050002), and the National Key Research and Development Program of China (No. 2022YFA1404203). J.Z. was supported by National Natural Science Foundation of China (No. 12401602).

%

\clearpage

\beginsupplement

\newpage

\pagenumbering{arabic} 

\begin{onecolumngrid}
\begin{center}
{\bf 
{\Large Supplementary Material for} \\  \vspace{0.2cm}
{\large ``Disorder-Induced Logarithmic Spreading and Localization of Quantum Fluctuations''}
}\\
\quad
\\
Ang Yang, Jingxuan Zhang, Lei Ying\\
\end{center}
\end{onecolumngrid}

\maketitle

\vspace{+1cm}

\begin{twocolumngrid}

\section{I. Time-dependent Bogoliubov-de Gennes description}

\subsection{A. Derivation of Bogoliubov-de Gennes equation}

Here, starting from the BH Hamiltonian in Eq.~\eqref{eq:Hintro}, we briefly review the Bogoliubov-de Gennes (BdG) equation, which describes the quasiparticle excitations in the larger-$N$ limit. Assume a macroscopically condensate wavefunction $\phi$ such that $\rho\lvert\phi\rangle=N_0\lvert\phi\rangle$ with $N_0\approx N$. We introduce the fluctuation operator $\delta\hat{b}, \delta \hat {b}^\dagger$ as
\begin{equation}
\begin{split}
\hat b_i &= \phi_i \hat{a}_{\phi} + \delta \hat b_i, \quad \hat{b}_i^\dagger = \phi_i^* \hat{a}^{\dagger}_{\phi} + \delta \hat{b}_i^\dagger, \\
\hat{a}_{\phi}&=\sum_i \phi_i^* \hat{b}_i,
\qquad \hat{N}=\hat{a}^{\dagger}_{\phi}\hat{a}_{\phi}+\sum_i \delta \hat{b}_i^\dagger\delta \hat b_i,
\end{split}
\end{equation}
where the fluctuation mode is orthogonal to the condensate mode that $\sum_i \phi_i^* \delta\hat{b}_i=0$. Then the commutation relations of $\hat{a}_{\phi}$ and $\delta\hat{b}$ satisfy respectively
\begin{equation}
    \big[\hat{a}_{\phi},\hat{a}^{\dagger}_{\phi}\big]=1,\quad\big[\delta\hat{b}_i,\delta\hat{b}^{\dagger}_j\big]=P_{ij},\quad P_{ij}=\delta_{ij}-\phi_i\phi_j^*.
\end{equation}
After substituting the above equations into Eq.~\eqref{eq:Hintro} and expanding Eq.~\eqref{eq:Hintro} up to the second order of $\delta\hat{b}$, we have
\begin{equation}
    H = H^{(0)}+H^{(1)}+H^{(2)}+\cdots. 
\end{equation}
In detail, we have
\begin{equation}
\begin{aligned}
H^{(0)} &= N\Big[-J \sum_i (\phi_i^* \phi_{i+1} + \phi_{i+1}^* \phi_i)
+ \sum_i \lambda_i |\phi_i|^2
+ \frac{U}{2} \sum_i |\phi_i|^4\Big], \\
H^{(1)} &= \sqrt{N}\sum_i \Big[ \delta \hat b_i^\dagger F_i(\phi) + \delta \hat b_i F_i^*(\phi) \Big], \\
F_i(\phi) &= -J (\phi_{i-1} + \phi_{i+1}) + \lambda_i \phi_i + U |\phi_i|^2 \phi_i,
\end{aligned}
\end{equation}
where we have considered the classical field approximation that $\hat{a}_{\phi}, \hat{a}_{\phi}^{\dagger}\sim \sqrt{N}$. Particularly, when the condensate wavefunction $\phi_i$ is the static state of the GP equation, $H^{(1)}$ vanishes. The second-order term reads
\begin{equation}
\begin{split}
&H^{(2)} = -J \sum_i \left( \delta \hat b_i^\dagger \delta \hat b_{i+1} + \delta \hat b_{i+1}^\dagger \delta \hat b_i \right) 
+ \sum_i \lambda_i \delta \hat b_i^\dagger \delta \hat b_i \\
&+ \frac{U}{2N} \sum_i \Big( 4 |\phi_i|^2 \hat{a}^{\dagger}_{\phi}\hat{a}_{\phi}\delta \hat b_i^\dagger \delta \hat b_i + \phi_i^2\hat{a}_{\phi}\hat{a}_{\phi} \delta \hat b_i^\dagger \delta \hat b_i^\dagger + {\phi_i^*}^2\hat{a}^{\dagger}_{\phi}\hat{a}^{\dagger}_{\phi} \delta \hat b_i \delta \hat b_i \Big).
\end{split}
\end{equation}
To guarantee the particle-number conservation, we introduce new fluctuation operators:
\begin{equation}
\begin{split}
    \hat{\Lambda}_i&=\hat{\alpha}^{\dagger}_{\phi}\delta\hat{b}_i,\quad \hat{\alpha}^{\dagger}_{\phi}=\frac{\hat{a}^{\dagger}_{\phi}}{\sqrt{\hat{N}_\phi}}, \\
    \hat{\Lambda}^{\dagger}_i&=\delta\hat{b}^{\dagger}_i\hat{\alpha}_{\phi},\quad \hat{\alpha}_{\phi}=\frac{\hat{a}_{\phi}}{\sqrt{\hat{N}_\phi}},
\end{split}
\end{equation}
which means that the creation of a excited particle is accompanied by the annihilation of a condensate particle. One can check that $[\Lambda_i,\hat{N}]=0$ holds and get the commutation relations
\begin{equation}
    \left[\hat{\alpha}_{\phi},\hat{\alpha}_{\phi}^{\dagger}\right]=\sqrt{1-\frac{1}{\hat{N}_\phi}}-1,\quad 
    \left[\hat{\Lambda}_i,\hat{\Lambda}_j\right]=P_{ij}-\delta\hat{b}^{\dagger}_j\delta\hat{b}_i  \left[\hat{\alpha}_{\phi},\hat{\alpha}_{\phi}^{\dagger}\right].
\end{equation}
In the large particle limit $\hat{N}_\phi\sim N\gg1$ with small depletion, we have $[\hat{\Lambda}_i,\hat{\Lambda}_j]\approx P_{ij}$. Introducing the vectors: 
\begin{equation}
\hat{\bm{\Lambda}} =  \big[\hat{\Lambda}_1,  \cdots,   \hat{\Lambda}_L\big]^T , \qquad
\hat{\bm{\Lambda}}^\dagger = \big[\hat{\Lambda}_1^\dagger,  \cdots,   \hat{\Lambda}_L^\dagger\big]^T  ,
\end{equation}
then the quadratic Hamiltonian turns into
\begin{equation}
\hat H^{(2)} =
 (\hat{\bm{\Lambda}}^\dagger)^T \, \mathbb{A} \, \hat{\bm{\Lambda}}
+ \frac{1}{2} \Big[
     (\hat{\bm{\Lambda}}^\dagger)^T \, \mathbb{B} \,  (\hat{\bm{\Lambda}}^\dagger)^T
  +  \hat{\bm{\Lambda}} \, \mathbb{B}^* \,  \hat{\bm{\Lambda}}
\Big],
\end{equation}
with 
\begin{align}
\mathbb{A}_{ij} &= -J (\delta_{i,j+1} + \delta_{i,j-1}) + (\lambda_i + 2 U |\phi_i|^2) \delta_{ij}, \\
\mathbb{B}_{ij} &= U \phi_i^2 \delta_{ij}.
\end{align}

Here, we adopt the formalism in Refs.~\cite{castin1997instability,castin1998bdg} to examine $\frac{d}{dt}\hat{\Lambda}_i$ and consider the expansion of $\phi$ and $\hat{\Lambda}$ in powers of $1/\sqrt{N}$. In our case, the condensate wavefunction $\psi_i(t)$ evolves according to the GP equation, which can be considered as the lowest-order approximation of $\phi$. Then we arrive the time-dependent BdG equation of the fluctuation operators
\begin{equation}
\begin{split}
i \frac{d}{dt} 
\begin{pmatrix}
    \hat{\bm{\Lambda} }\\
    \hat{\bm{\Lambda}}^{\dagger}
\end{pmatrix}
= \mathcal{H}
\begin{pmatrix}
    \hat{\bm{\Lambda}} \\
    \hat{\bm{\Lambda}}^{\dagger}
\end{pmatrix}, \quad 
\mathcal{H}=\begin{pmatrix}
    \mathcal{A}(t) & \mathcal{B}(t) \\
    -\mathcal{B}(t)^* & -\mathcal{A}(t)^*
\end{pmatrix}, 
\end{split}
\end{equation}
with 
\begin{equation}
\begin{split}
\mathcal{A}(t)&=\hat{H}_{\mathrm{GP}}(t)+P(t)\mathrm{diag}[U\lvert\psi_i(t)\rvert^2]P(t),\\
\mathcal{B}(t)&=P(t)\mathrm{diag}[U\psi^2_i(t)]P^*(t),
\end{split}
\end{equation}
where $\hat{H}_{\mathrm{GP},ij}(t)=-J (\delta_{i,j+1} + \delta_{i,j-1}) + (\lambda_i + U |\psi_i(t)|^2) \delta_{ij}$ and $P_{ij}(t)=\delta_{ij}-\psi_i(t)\psi_j(t)^*$ guarantees the orthogonality between the condensate and the fluctuation modes. As the initial condition is $[\langle \bm{\Lambda}\rangle,\ \langle\bm{\Lambda}^\dagger\rangle]^T=[\bm{0},\ \bm{0}]^T$, the first-order contribution $\langle\hat{\Lambda}_j\rangle$ is always zero. To evaluate the second-order contribution, one can introduce the covariance matrix
\begin{equation}
{\mathcal{C}}(t) 
=\begin{pmatrix}
\langle  \hat \Lambda_i \hat \Lambda_j^\dagger \rangle & \langle  \hat \Lambda_i  \hat \Lambda_j \rangle \\
\langle \hat \Lambda_i^\dagger  \hat \Lambda_j^\dagger \rangle & \langle  \hat \Lambda_i^\dagger  \hat \Lambda_j \rangle
\end{pmatrix}_{2L \times 2L},
\end{equation}
which evolves according to 
\begin{equation}
i \frac{d}{dt} \mathcal{C}(t) = \mathcal{H}(t) \, \mathcal{C}(t) - \mathcal{C}(t) \, \mathcal{H}^{\dagger}(t),
\end{equation}
with the initial condition $\mathcal{C}(0)=[
\bm{P} \ \bm{0} ;\
\bm{0} \ \bm{0}]$ for the initial condensate many-body state. The quantum fluctuation we define in terms of the particle-number operator $\hat{n}_j$ reads
\begin{equation}
\begin{split}
    C(j,t) &=\left\lvert\frac{\langle\hat{b}_j^{\dagger}(t)\hat{b}_j(t)\rangle}{N}-\lvert\psi_j(t)\rvert^2\right\rvert \\
    &\approx\left\lvert \frac{(N-\delta N)\lvert\psi_j(t)\rvert^2+\langle\hat{\Lambda}_j^{\dagger}(t)\hat{\Lambda}_j(t)\rangle}{N}- \lvert\psi_j(t)\rvert^2\right\rvert \\
    &\approx\left\lvert\frac{\langle\hat{\Lambda}_j^{\dagger}(t)\hat{\Lambda}_j(t)\rangle}{N}\right\rvert,
\end{split}
\end{equation}
where $\delta N$ is the noncondensed particle number and we have considered the large-$N$ limit, thus $\langle\hat{\Lambda}^{\dagger}_j\hat{\Lambda}_j\rangle\approx\langle\delta\hat{b}_j^{\dagger}\delta\hat{b}_j\rangle$ and $\langle\hat{\Lambda}_j\rangle\approx\langle\delta\hat{b}_j\rangle$. Therefore, the quantum fluctuation is dominated by the second-order contribution. Note that the BdG description is mostly valid at short times when the depletion remains small, i.e., $\sum_i\langle  \hat \Lambda_i^\dagger  \hat \Lambda_i \rangle\ll N$. In Fig.~\ref{transport}, using the BdG equation, we show the evolution of the spatial profile of $\overline{C(j,t)}$. Even though they differ in magnitude, the BdG results properly capture the early time profile of finite-N results. 
In Fig.~\ref{bdg_transport}, we also show the dynamics of RMSD and the depletion for different $\lambda$ and disorder types. Except for the sudden growth at $t=0.1$ due to the initial condition, the BdG results closely follow the finite-$N$ results. Particularly, at intermediate times, RMSD dynamics reproduce a logarithmic growth and it slows down as $\lambda$ increases. This is attributed to the fact that the RMSD dynamics only depend on the profile, not the overall magnitude. The fast growth of depletion after $t\sim10$ also indicates the validity of BdG description at short times. Figure~\ref{bdg_projector} also demonstrates the effect of the projector $P(t)$, which is not considered in the usual symmetry-breaking BdG approach. 

\begin{figure}
\centering
\includegraphics[width=\linewidth]{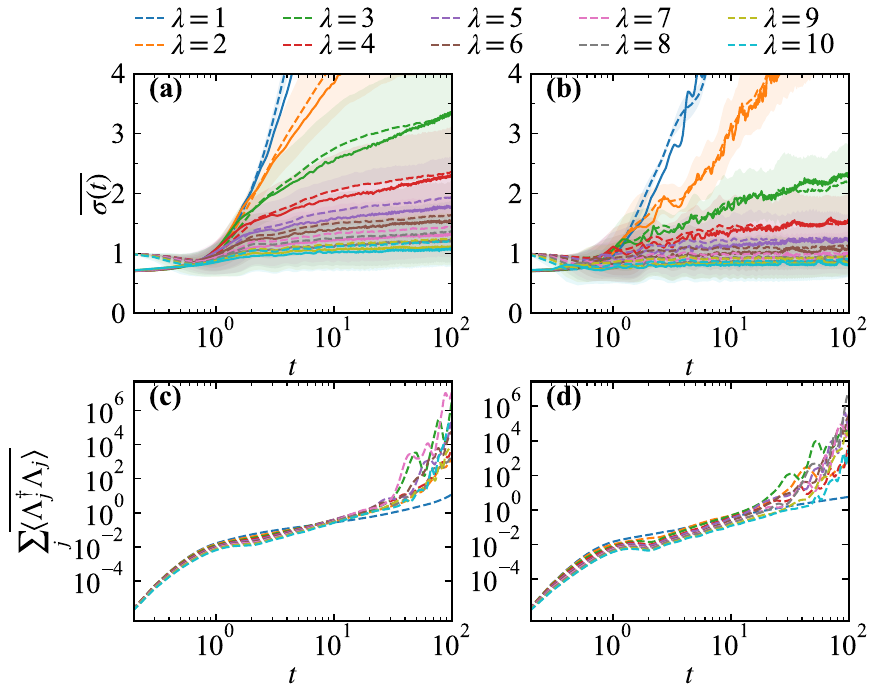}
\caption{{\bf Comparison between BdG and finite-N results}. The averaged RMSD (a) $\overline{\sigma(t)}$ and depletion (c) $\sum_j\overline{\langle\Lambda^{\dagger}_j\Lambda_j\rangle}$ as a function of time $t$ for different disorder strengths $\lambda$ in a random setting. (b) and (d): Similar data as in (a) and (c) but in a quasiperiodic setting. The color shaded area denotes the standard deviation at each disorder strength. The system size, and the sampling number are respectively $L=100$ and $N_{\mathrm{s}}=1000$. The solid lines are from $N=4$ for comparison with BdG results (dashed lines).
}
\label{bdg_transport}
\end{figure}

\begin{figure}
\centering
\includegraphics[width=\linewidth]{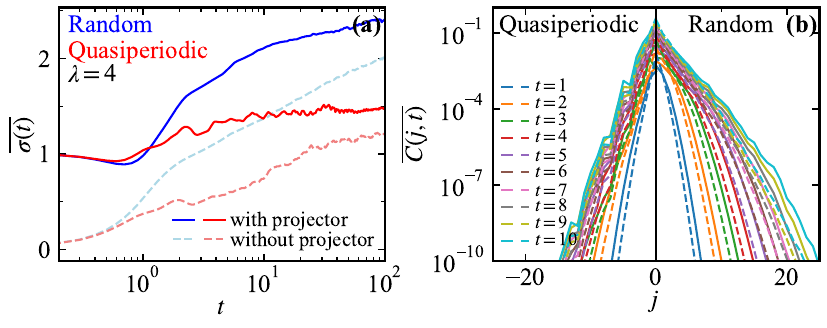}
\caption{{\bf Effect of projector}. (a) The averaged RMSD dynamics $\overline{\sigma(t)}$ for different disorder settings at $\lambda=4$. The dashed lines denote the BdG results with ${P}={I}$ during evolution. (b) The corresponding spatial profile of $\overline{C(j,t)}$ at different times are compared. The system size, and the sampling number are respectively $L=100$ and $N_{\mathrm{s}}=1000$.
}
\label{bdg_projector}
\end{figure}

\begin{figure}
\centering
\includegraphics[width=\linewidth]{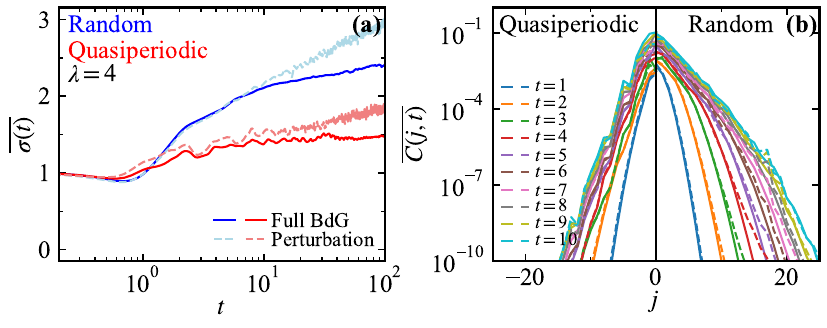}
\caption{{\bf Comparison between BdG and perturbative results}. (a) The averaged RMSD dynamics $\overline{\sigma(t)}$ for different disorder settings at $\lambda=4$. (b) The corresponding spatial profile of $\overline{C(j,t)}$ at different times are compared. For BdG results (solid lines), the system size and the sampling number are respectively $L=100$ and $N_{\mathrm{s}}=1000$; for perturbative results (dashed lines), $L=100$ and $N_{\mathrm{s}}=100$.
}
\label{perturbation_comparison}
\end{figure}

\begin{figure*}
\centering
\includegraphics[width=0.9\linewidth]{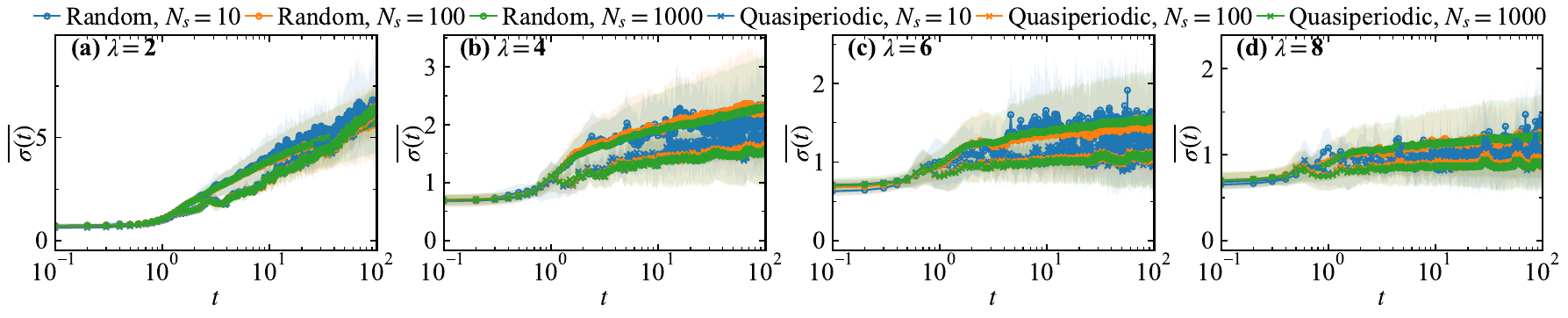}
\caption{
(a-d) The averaged RMSD dynamics for different sampling numbers $N_{\mathrm{s}}$ and disorder strengths $\lambda$, in two disorder settings. The color-shaded area denotes the corresponding standard deviation. The particle number and the system size are $N=4$ and $L=50$, respectively.
}
\label{sampling_number}
\end{figure*}

\begin{figure*}
\centering
\includegraphics[width=0.75\linewidth]{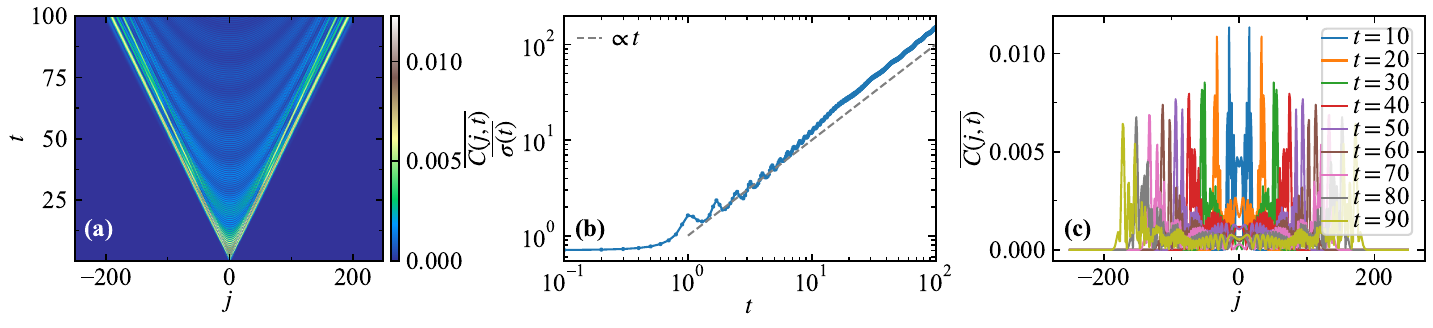}
\caption{
(a) The averaged quantum fluctuation density distribution $\overline{C(j,t)}$ dynamics for the disorder-free case. (b) The corresponding RSMD dynamics, and the grey-dashed line denotes ballistic transport $\propto t$. (c) The corresponding profiles of $\overline{C(j,t)}$ at different times. The particle number and the system size are $N=2$ and $L=500$, respectively.
}
\label{no_disorder}
\end{figure*}

\begin{figure*}
\centering
\includegraphics[width=0.95\linewidth]{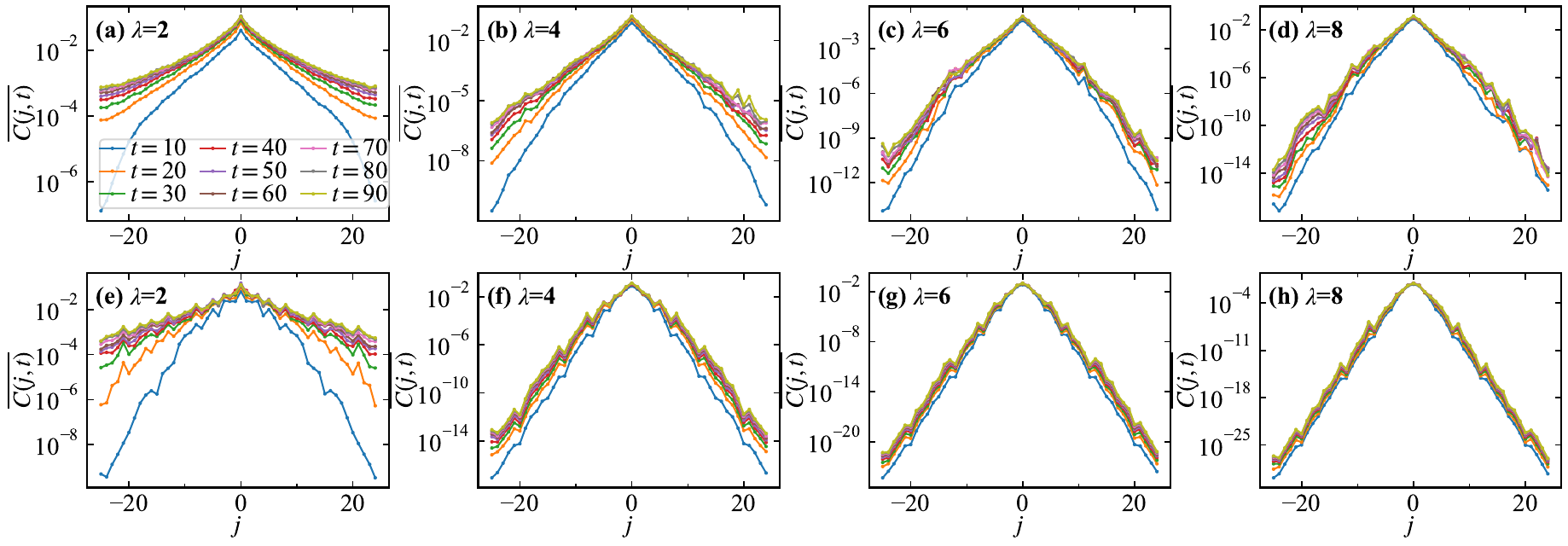}
\caption{
(a-d) The averaged profiles of quantum fluctuation density $\overline{C(j,t)}$ at different times and disorder strengths, in a random setting. (e-h) The similar data as in (a-d) but in a quasiperiodic setting. The particle number and the system size are $N=4$ and $L=50$, respectively.
}
\label{profile_evolution}
\end{figure*}

\begin{figure*}
\centering
\includegraphics[width=\linewidth]{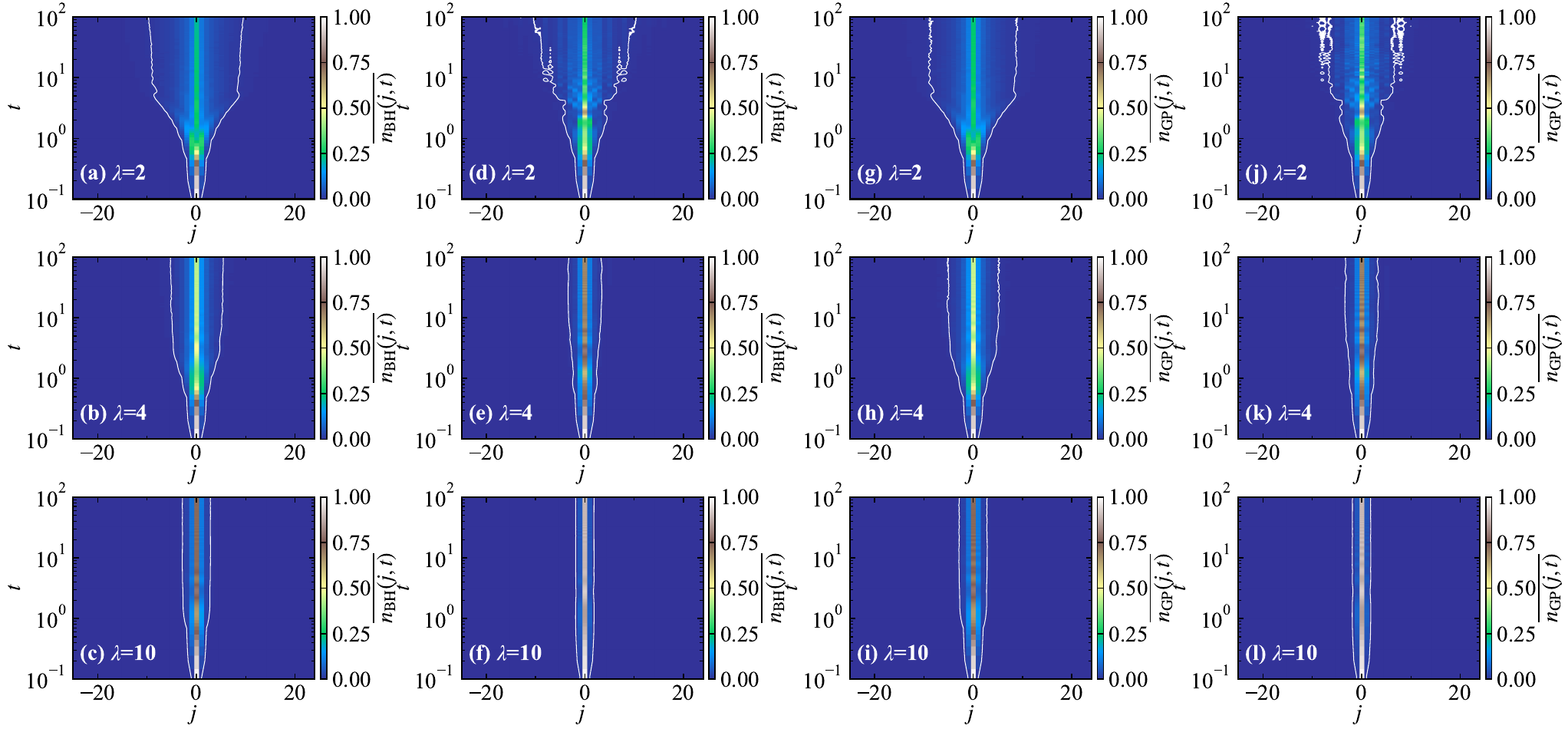}
\caption{
(a-c) The averaged particle density  $\overline{n_{\mathrm{BH}}(j,t)}$ dynamics in the BH model, for different disorder strengths in a random setting. (d-f) Similar data as in (a-c) but in a quasiperiodic setting. (g-i) The averaged particle density  $\overline{n_{\mathrm{GP}}(j,t)}$ dynamics in the GP model, for different disorder strengths in a random setting. (j-l) Similar data as in (g-i) but in a quasiperiodic setting. The white lines denote the contour at $10^{-2}$. The particle number and the system size are $N=4$ and $L=50$, respectively.
}

\label{density_evolution}
\end{figure*}
\begin{figure*}
\centering
\includegraphics[width=0.9\linewidth]{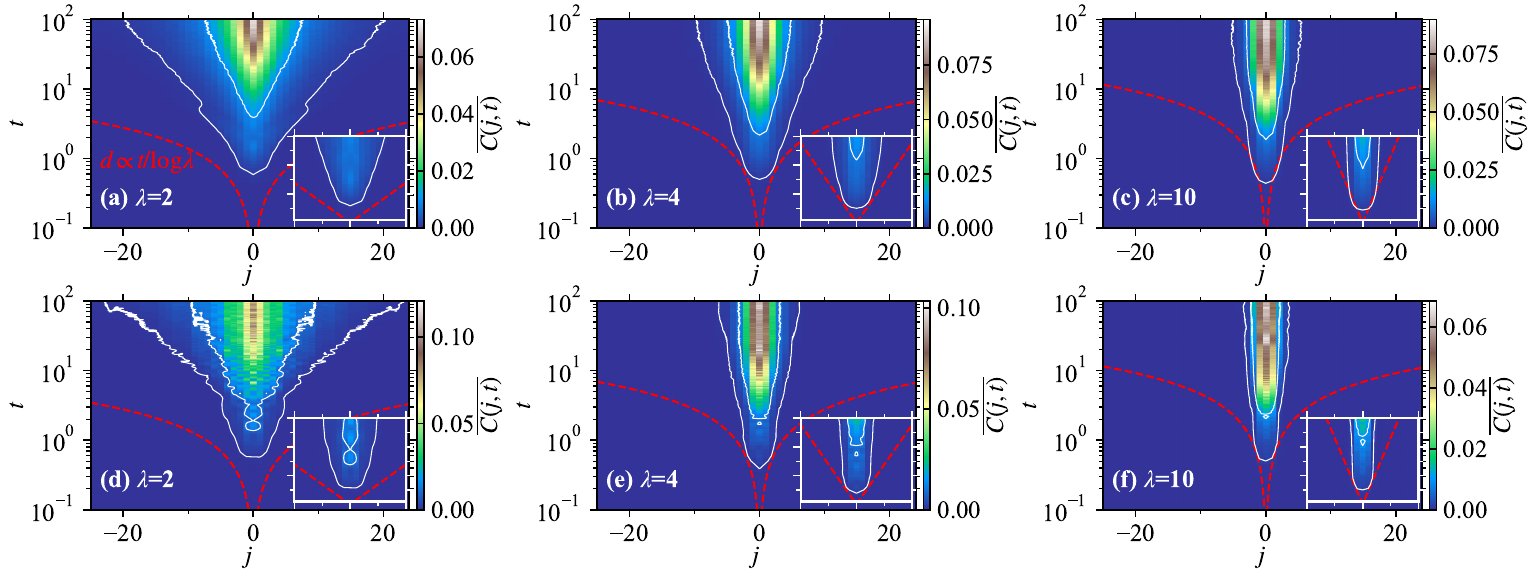}
\caption{
{\bf Another initial state.} (a-c) The averaged quantum fluctuation density distribution $\overline{C(j,t)}$ as a function of lattice index $j$ and time $t$ for different random disorder strengths $\lambda$. (d-f) The similar data as in (a-c) but in a quasiperiodic disorder setting. Insets show the detailed spatial-temporal distributions of $\overline{C(j,t)}$ at early times in a linear scale. The white lines denote the contour at $\overline{C(j,t)}=10^{-3}$ and $10^{-2}$, while the red lines denote the estimated ballistic bound of transport $d\propto t/\mathrm{log}\lambda$. The particle number and the system size are $N=4$ and $L=50$, respectively. The initial state is $\psi(t=0)=\frac{1}{\sqrt{3}}(\delta_{j_0-1}+\delta_{j_0}+\delta_{j_0+1})$.
}

\label{initial_state2_transport}
\end{figure*}

\begin{figure}
\centering
\includegraphics[width=1\linewidth]{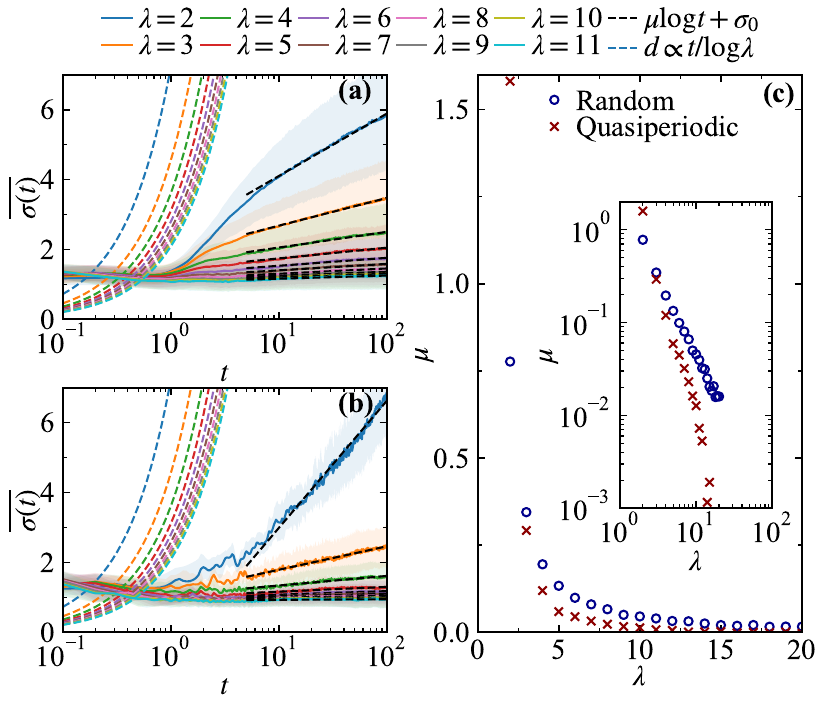}
\caption{
{\bf Another initial state.} (a-b) The averaged RMSD $\overline{\sigma(t)}$ as a function of time $t$, for different disorder strengths $\lambda$ in a random (a) and quasiperiodic (b) setting. The black dashed lines denote the fitting curve $\mu\mathrm{log}t+\sigma_0$. The colored lines denote the estimate ballistic bound of transport $d\propto t/\mathrm{log}\lambda$, and the color-shaded area denotes the standard deviation at each disorder strength. (c) The fitted slope $\mu$ as a function of disorder strength $\lambda$ for random and quasiperiodic settings, and inset shows the same data in the log-log scale. The particle number and the system size are $N=4$ and $L=50$, respectively. The initial state is $\psi(t=0)=\frac{1}{\sqrt{3}}(\delta_{j_0-1}+\delta_{j_0}+\delta_{j_0+1})$.
}

\label{initial_state2_sigma}
\end{figure}

\begin{figure*}
\centering
\includegraphics[width=0.7\linewidth]{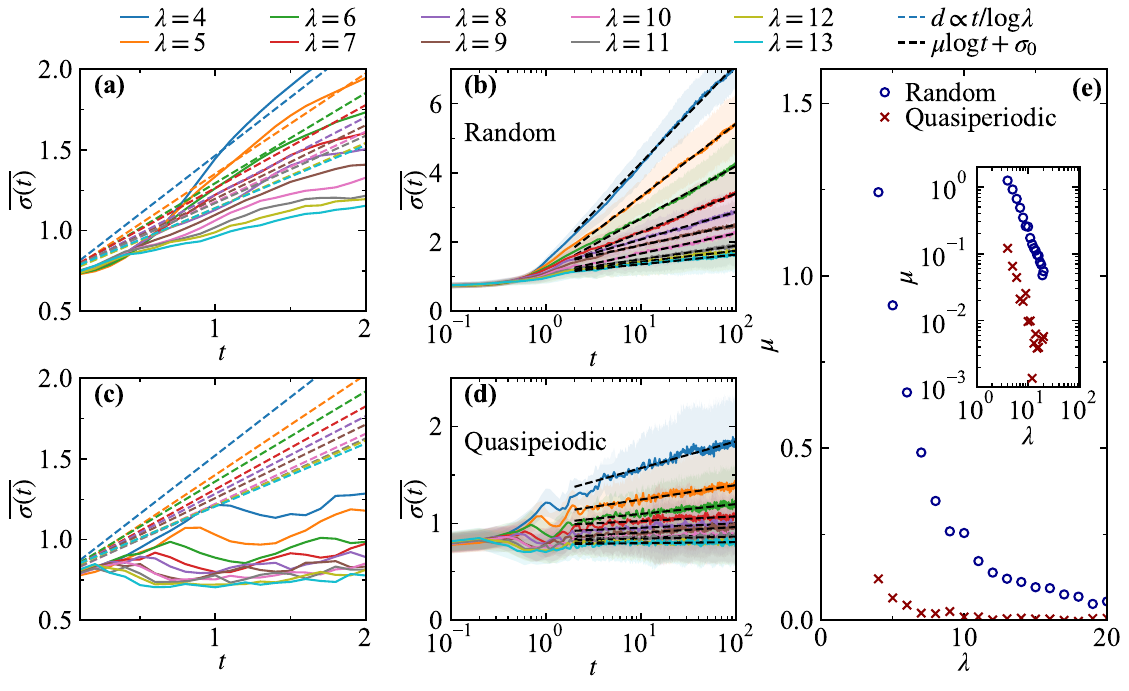}
\caption{{\bf 2D case}. (a-b) The averaged RMSD $\overline{\sigma(t)}$ as a function of time $t$ in linear (a) and semilog scale (b), for different random disorder strengths $\lambda$. (c-d) The similar data as in (a-b) but in a quasiperiodic disorder setting. The black dashed lines denote the fitting curve $\mu\mathrm{log}t+\sigma_0$. The color dashed lines denote the estimate ballistic bound of transport $d\propto t/\mathrm{log}\lambda$, and the color shaded area denotes the standard deviation at each disorder strength. (e) The fitted slope $\mu$ as a function of disorder strength $\lambda$ for random and quasiperiodic settings, and inset shows the same data in the log-log scale. The particle number, the system size, and the sampling number are respectively $N=2$ and $L_x=L_y=25$, and $N_{\mathrm{s}}=100$.
}
\label{2D_case}
\end{figure*}

\begin{figure}
\centering
\includegraphics[width=1\linewidth]{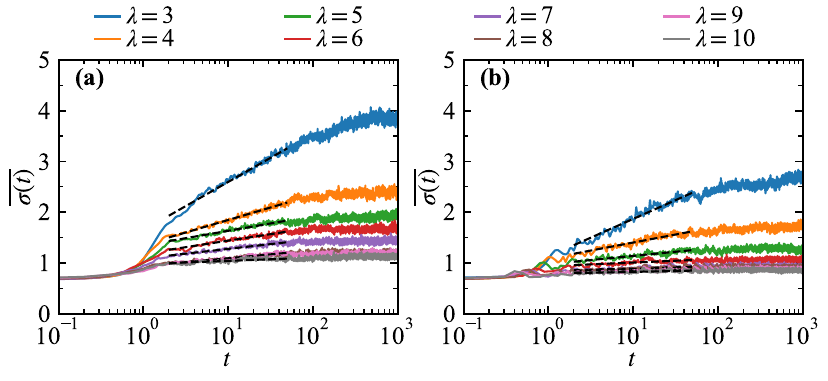}
\caption{
The averaged RMSD $\overline{\sigma(t)}$ as a function of time $t$ for different disorder strengths $\lambda$ in a random (a) and quasiperiodic (b) setting. The black dashed lines denote the fitting curve $\mu\mathrm{log}t+\sigma_0$. The particle number, the system size, and the sampling number are respectively $N=2$ and $L=150$, and $N_{\mathrm{s}}=100$.
}

\label{long_time}
\end{figure}

\begin{figure}
\centering
\includegraphics[width=1\linewidth]{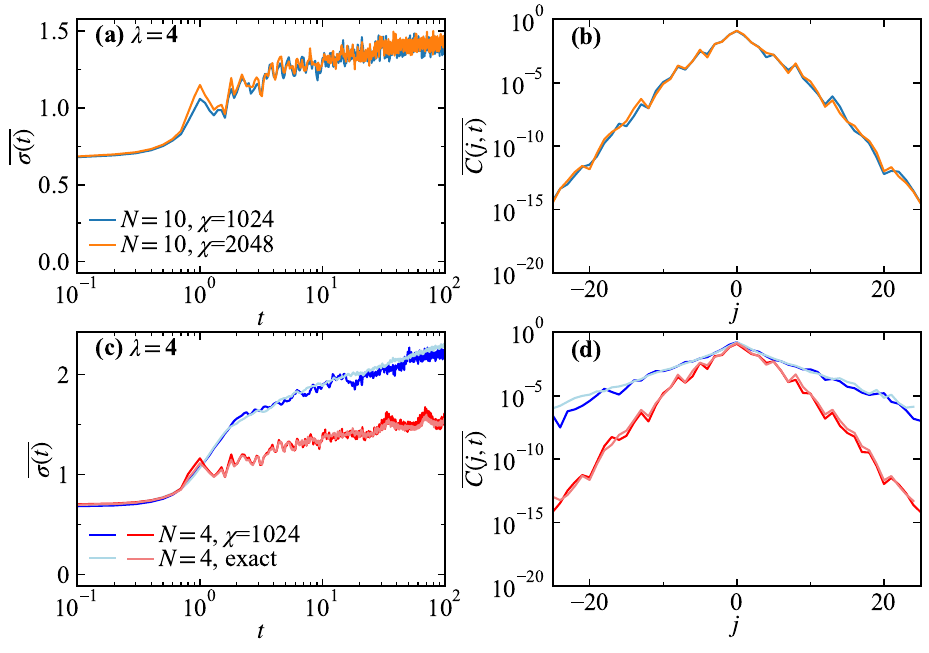}
\caption{{\bf Convergence of TEBD results}. (a-b) The RMSD dynamics (a) and the spatial profile of quantum fluctuation $\overline{C(j,t)}$ at $t=100$ (b) for different bond dimension $\chi$, at a quasiperiodic disorder strength $\lambda=4$ and $N=10$. (c-d) Similar as in (a-b), but the TEBD and exact results are compared for random (blue) and quasiperiodic (red) settings at $N=4$. The system size and sampling number for the TEBD results are $L=100$ (a-b), $L=500$ (c-d) and $N_{\mathrm{s}}=100$.
}
\label{tebd}
\end{figure}

\subsection{B. Perturbative analysis of logarithmic spreading of quantum fluctuation}
\emph{BdG dynamics and localized eigenmodes---}We investigate the spreading of quantum fluctuations in a weakly interacting disordered bosonic system using the BdG framework. We approximate the projector $P(t)$  by its instantaneous form and treat the time-dependent anomalous pairing interaction as a weak perturbation to the normal BdG dynamics. The corresponding BdG Hamiltonian reads
\begin{equation}
\mathcal{H}
=
\begin{pmatrix}
\mathcal A & \mathcal B(t)\\
-\mathcal B^*(t) & -\mathcal A^*
\end{pmatrix}.
\label{eq:bdg_dynamics}
\end{equation}
We assume that the GP wavefunction remains spatially localized around the initially occupied site $j_0$, with an approximately stationary density profile but site-dependent phases,
\begin{equation}
\psi_j(t)=\sqrt{\rho_j}\,e^{-i\theta_j(t)},
\quad\mathrm{with}\quad
\sqrt{\rho_j}\sim e^{-|j-j_0|/\xi_{\mathrm{GP}}},
\label{eq:gp_localization}
\end{equation}
where $\xi_{\mathrm{GP}}$ characterizes the localization length of the GP wavefunction. Under this approximation, the normal BdG matrix $\mathcal A$ is treated as time-independent, whereas the anomalous pairing matrix retains the temporal evolution of the condensate phases,
\begin{equation}
\mathcal B_{jl}(t)=U\rho_j e^{-2i\theta_j(t)}\delta_{jl}.
\label{eq:pairing_matrix}
\end{equation}
We further assume that the eigenfunctions of the normal BdG matrix are exponentially localized. Specifically,
\begin{equation}
\mathcal A u_\alpha=\varepsilon_\alpha u_\alpha,
\label{eq:normal_eigenmodes}
\end{equation}
with
\begin{equation}
|u_\alpha(j)|\sim e^{-|j-x_\alpha|/\xi_A},
\label{eq:eigenmode_localization}
\end{equation}
where $x_\alpha$ is the localization center of eigenmode $\alpha$ and $\xi_A$ denotes its characteristic localization length.

Expanding the fluctuation operator in the eigenbasis of $\mathcal A$,
\begin{equation}
\hat{\Lambda}_j(t)=\sum_\alpha u_\alpha(j)\hat{c}_\alpha(t), \qquad
\hat{\Lambda}^\dagger_j(t)=\sum_\alpha u^\ast_\alpha(j)\hat{c}_\alpha^\dagger(t),
\label{eq:mode_expansion}
\end{equation}
we obtain the equations of motion
\begin{equation}
\begin{split}
i\frac{d\hat{c}_\alpha}{dt}
=&
\varepsilon_\alpha\hat{c}_\alpha
+
\sum_\beta g_{\alpha\beta}(t)\hat{c}_\beta^\dagger, \\
i\frac{d\hat{c}_\alpha^\dagger}{dt}
=&
-\varepsilon_\alpha\hat{c}_\alpha^\dagger
-
\sum_\beta g_{\alpha\beta}^*(t)\hat{c}_\beta,
\end{split}
\label{eq:conjugate_mode_dynamics}
\end{equation}
where the pairing matrix elements are
\begin{equation}
g_{\alpha\beta}(t)
=
U\sum_j u_\alpha^*(j)u_\beta^*(j)\rho_j e^{-2i\theta_j(t)}.
\label{eq:mode_dynamics}
\end{equation}
The exponential localization of the GP wavefunction and the normal eigenmodes suppresses the pairing interaction between spatially separated modes. Consequently, the excitation of distant localized modes is expected to occur on progressively longer timescales.

\emph{Weak pairing and disorder-averaged excitation dynamics---}Since the anomalous pairing interaction is weak, we treat $g_{\alpha\beta}(t)$ perturbatively. To first order, the anomalous pairing amplitude is
\begin{equation}
Y_{\alpha\beta}(t)
=
-ie^{-i\varepsilon_\alpha t}
\int_0^t ds\,
g_{\alpha\beta}(s)e^{i(\varepsilon_\alpha+\varepsilon_\beta)s}.
\label{eq:anomalous_amplitude}
\end{equation}
For an initial fluctuation vacuum, the local fluctuation occupation is denoted by $\Pi_j(t)$ and can be expressed as
\begin{equation}
\Pi_j(t)=\left\langle \hat\Lambda_j^\dagger(t)\hat\Lambda_j(t) \right\rangle = 
\sum_\beta
\left|
\sum_\alpha u_\alpha(j)Y_{\alpha\beta}(t)
\right|^2.
\label{eq:local_fluctuation}
\end{equation}
In Fig.~\ref{perturbation_comparison}, we show the accuracy of the above perturbative formula, compared to the full BdG results.

We now consider the disorder-averaged fluctuation occupation. Assuming that interference contributions between different localized modes $\alpha\neq\alpha'$ are suppressed after averaging over disorder realizations, we obtain
\begin{equation}
\overline{\Pi_j(t)}
\approx
\sum_{\alpha,\beta}
\overline{
|u_\alpha(j)|^2
|Y_{\alpha\beta}(t)|^2
}.
\label{eq:incoherent_fluctuation}
\end{equation}

To estimate the temporal accumulation of pairing-generated excitations, we further approximate the local GP phases as $\theta_l(t)\approx\omega_lt+\theta_l(0)$, where $\omega_l$ denotes an effective local phase frequency. The corresponding frequency mismatch for a pairing process involving modes $\alpha$ and $\beta$ at site $l$ is
\begin{equation}
\Delta_{\alpha\beta,l}
=
\varepsilon_\alpha+\varepsilon_\beta-2\omega_l.
\label{eq:frequency_mismatch}
\end{equation}
Using time integral $\int_0^t e^{i\Delta_{\alpha\beta,l}s}ds=te^{i\Delta_{\alpha\beta,l}t/2}\mathrm{sinc}(\Delta_{\alpha\beta,l}t/2)$ in $Y_{\alpha\beta}(t)$, it reads
\begin{equation}
{\begin{aligned}
Y_{\alpha\beta}(t)=-iUt\,e^{-i\epsilon_\alpha t}
\sum_l u_\alpha^*(l)u_\beta^*(l)\rho_l e^{i\Delta_{\alpha\beta,l}t/2}
\operatorname{sinc}\!\left(\frac{\Delta_{\alpha\beta,l}t}{2}\right),
\end{aligned}}
\label{eq:Yexplicit}
\end{equation}
and thus its modulus square reads
\begin{align}
 \overline{|Y_{\alpha\beta}(t)|^2}
 &\approx U^2t^2\,\overline{\sum_l |u_\alpha(l)|^2|u_\beta(l)|^2\rho_l^2\,\mathrm{sinc}^2\left(\frac{\Delta_{\alpha\beta,l}t}{2}\right)},
 \label{eq:avgY}
\end{align}
where we further neglects interference between distinct pairing-source sites $l\neq l'$. By defining the overlap weight
\begin{equation}
\mathcal W_{\alpha\beta l}(j)=|u_\alpha(j)|^2|u_\alpha(l)|^2 |u_\beta(l)|^2\rho_l^2,
\end{equation}
the disorder-averaged occupation becomes
\begin{align}
\overline{\Pi_j(t)}
&\approx U^2t^2\,\overline{\sum_{\alpha,\beta,l}
\mathcal W_{\alpha\beta l}(j)
\operatorname{sinc}^2\!\left(\frac{\Delta_{\alpha\beta,l}t}{2}\right)}.
\label{eq:incoherent_sources}
\end{align}
Introducing the disorder-averaged, overlap-weighted pairing spectral density
\begin{equation}
{S_j(\Delta)=U^2\,\overline{\sum_{\alpha,\beta,l}
\mathcal W_{\alpha\beta l}(j)\delta(\Delta-\Delta_{\alpha\beta,l})},}
\label{eq:spectral}
\end{equation}
we obtain
\begin{equation}
{\overline{\Pi_j(t)}\approx t^2\int d\Delta\,S_j(\Delta)\,\mathrm{sinc}^2\left(\frac{\Delta t}{2}\right).}
 \label{eq:disavgPi}
\end{equation}
For times shorter than the inverse detuning width $W^{-1}$, Eq.~\eqref{eq:disavgPi} reproduces the initial coherent quadratic growth, $\overline{\Pi_j(t)}\approx t^2\int d\Delta\,S_j(\Delta)$ when $Wt\ll1$. For $Wt\gg1$, if $S_j(\Delta)$ is smooth and nonzero at resonance, $t\,\mathrm{sinc}^2(\Delta t/2)\to2\pi\delta(\Delta)$ in the distributional sense. Hence
\begin{equation}
{\overline{\Pi_j(t)}\approx\Gamma_jt,\qquad\Gamma_j=2\pi S_j(0),\qquad Wt\gg1,}
 \label{eq:golden}
\end{equation}

Since the condensate and normal eigenfunctions are exponentially localized, the spatial overlap associated with pairing processes involving distant sites becomes exponentially small. Assuming that the resonant spectral weight preserves this exponential spatial dependence, we estimate
\begin{equation}
\Gamma_j
\approx
\Gamma_0
e^{-|j-j_0|/\ell_{\mathrm{eff}}},
\label{eq:spatial_pairing_rate}
\end{equation}
where $\ell_{\mathrm{eff}}$ is the effective decay length of the disorder-averaged pairing rate. This length depends on the localization properties of the GP wavefunction, the normal BdG eigenmodes, and the distribution of relevant resonances.
By defining the excitation distance $d_\eta(t)$ through a fixed fluctuation occupation $\overline{\Pi_{j_0\pm d(t)}(t)}=\eta$, we obtain a logarithmic fluctuation front 
\begin{equation}
  d_\mathrm{\eta}(t)\approx\ell_\mathrm{eff}\log\left({\Gamma_0t}/{\eta}\right).
\end{equation}
If the GP density or pairing correlations themselves evolve appreciably, a time-dependent $\xi_{\mathrm{GP}}(t)$ cannot be merely substituted into their stationary spectral density. In that case, the more general disorder-averaged result, still neglecting $\alpha\neq\alpha'$ mode interference, involves the two-time pairing correlator:
\begin{equation}
\begin{split}
\overline{\Pi_j(t)}\approx \sum_{\alpha,\beta}
\int_0^t ds\int_0^t ds'
\overline{|u_\alpha(j)|^2g_{\alpha\beta}(s)g_{\alpha\beta}^*(s')
 e^{i(\varepsilon_\alpha+\varepsilon_\beta)(s-s')}},
 \end{split}
\label{eq:two_time}
\end{equation}
which cannot give rise to an explicit form. Below, we introduce a phenomenological kinetic description to account for the time-dependent growth of localization length $\xi_{\mathrm{GP}}(t)$.

\subsection{C. Phenomenological description of quantum fluctuation}

\emph{Time-dependent spatial suppression and local saturation---}
For a state initially concentrated around $j_0$, nearest-neighbor hopping gives the short-time propagator $K_{j_0+d,j_0}(t) \approx (iJt)^d/d!$. This strongly suppresses the early development of distant fluctuations. We represent this evolving spatial suppression by a time-dependent development length $\xi(t)$, with $\xi(0)=0$ and $\xi(t)\to\xi_\infty>0$. The simplest smooth interpolation is
\begin{equation}
\xi(t)=\xi_\infty\left(1-e^{-t/t_c}\right).
 \label{eq:ell}
\end{equation}
To describe the eventual finite spatial width, we further assume a stationary exponential occupation profile $\Pi_{\rm sat}(d)=\Pi_\infty e^{-d/\ell_\mathrm{f}}$, where $d=|j-j_0|$. Following the disorder-averaged golden-rule rate in Eq.~\eqref{eq:golden}, we model the time-dependent, spatially suppressed coarse-grained generation rate as $\Gamma(d,t)=2\pi S_r(0;t)\approx\Gamma_0e^{-d/\ell_\mathrm{f}}e^{-d/\xi(t)}$. In its stationary limit, this rate has $\ell_{\mathrm{eff}}^{-1}=\ell_\mathrm{f}^{-1}+\xi_\infty^{-1}$. This factorization of the spectral weight into two length scales is an additional modeling assumption, not implied by Eq.~\eqref{eq:spectral}. 

A minimal local-saturation closure is then $\partial_t\Pi=\Gamma(d,t)[1-\Pi/\Pi_{\rm sat}(d)]$, or equivalently
\begin{equation}
 \frac{\partial\Pi(d,t)}{\partial t}=\frac{\Pi_{\rm sat}(d)-\Pi(d,t)}{\tau_0}\,e^{-d/\xi(t)},\qquad \tau_0=\frac{\Pi_\infty}{\Gamma_0}.
 \label{eq:kinetics}
\end{equation}
Here $\Gamma_0$ is the coarse-grained central generation rate. Both the specific function $\xi(t)$ and the saturation factor in Eq.~\eqref{eq:kinetics} are phenomenological assumptions informed by the disorder-averaged, localized BdG pairing mechanism; they do not follow as exact consequences of first-order BdG perturbation theory.

With $\Pi(d,0)=0$, Eq.~\eqref{eq:kinetics} has the exact solution
\begin{equation}
{\begin{aligned}
\Pi(d,t)&=\Pi_{\mathrm{sat}}(d)
\big[1-e^{-F(d,t)}\big],\\
F(d,t)&=\frac{1}{\tau_0}\int_0^t ds\,
 \exp\!\left[-\frac{d}{\xi(s)}\right].
\end{aligned}}
\label{eq:Pi}
\end{equation}
The limit $e^{-d/\xi(0)}=0$ is understood for $d>0$. At short times, $\xi(s)\approx as$ with $a=\xi_\infty/t_c$, hence for $d>0$
\begin{equation}
\begin{aligned}
&F(d,t)\approx\frac{1}{\tau_0}\left[te^{-b/t}-bE_1(b/t)\right],\\
&E_1(x)=\int_x^\infty\frac{e^{-u}}{u}\,du
\quad\mathrm{and}\quad b=d/a.
\end{aligned}
\label{eq:shortF}
\end{equation}
In particular, $F(r,t)\sim t^2e^{-b/t}/(b\tau_0)$ when $t\ll b$, showing an intrinsic, strongly distance-dependent onset.

\emph{Recovery of the intermediate-time linear-growth profile---}
The disorder-averaged BdG result in Eq.~\eqref{eq:golden} predicts a linear occupation at a fixed site after detuning-induced dephasing and before perturbation theory breaks down. We now show explicitly that the effective model in Eq.~\eqref{eq:Pi} retains the same intermediate regime. In particular, the time dependence of $\xi(t)$ must be integrated first; simply replacing $\xi(s)$ by $\xi(t)$ at all earlier times would lose its onset history. For the choice in Eq.~\eqref{eq:ell}, we define $c=d/\xi_\infty$ and write the integral in Eq.~\eqref{eq:Pi} exactly as
\begin{equation}
\begin{aligned}
F(d,t)&=\frac{e^{-c}}{\tau_0}\left[t-\delta_t(d)\right],\\
\delta_t(d)&=\int_0^t ds\,
 \left[1-\exp\!\left(-\frac{c}{e^{s/t_c}-1}\right)\right].
\end{aligned}
\label{eq:time_shift}
\end{equation}
For any fixed $d>0$, the effective delay $\delta_t(d)$ approaches a finite constant at $t\gg t_c$; at $d=0$ it vanishes. The limiting shift can be expressed through the exponential integral as
\begin{equation}
 \delta_\infty(d)=t_c\left[\gamma_{\mathrm E}+\mathrm{log} c+e^cE_1(c)\right],\qquad c>0,
 \label{eq:delta_inf}
\end{equation}
where $\gamma_{\mathrm E}$ is Euler's constant. Consequently, when $t$ is large compared with the effective shift, $F(r,t)\approx (t/\tau_0)e^{-d/\xi_\infty}$. Now consider a site in the unsaturated part of the fluctuation distribution, for which $F(d,t)\ll1$. Expanding the exact solution of Eq.~\eqref{eq:Pi} gives
\begin{equation}
\begin{split}
\Pi(d,t)&\approx \Pi_{\mathrm{sat}}(d)F(d,t)\\
&\approx\frac{\Pi_\infty}{\tau_0}\big[t-\delta_\infty(d)\big]
 e^{-d/\ell_{\mathrm{eff}}}\\
&\approx{\Gamma_0t\,e^{-d/\ell_{\mathrm{eff}}}},
\end{split}\label{eq:intermediate_linear}
\end{equation}
where $\Gamma_0=\Pi_\infty/\tau_0$ and $\ell_{\mathrm{eff}}^{-1}=\ell_\mathrm{f}^{-1}+\xi_\infty^{-1}$.
Thus the effective model recovers precisely the disorder-averaged BdG form $\Pi(d,t)\sim t\,e^{-d/\ell_{\mathrm{eff}}}$, when the coarse-grained rate is approximately stationary. In this effective model, the complete evolution consists of the early coherent growth $\Pi\propto t^2$, subsequent disorder-averaged linear growth $\Pi\propto t e^{-d/\ell_{\mathrm{eff}}}$ where $F\ll1$, and eventual local saturation $\Pi\to\Pi_\infty e^{-d/\ell_\mathrm{f}}$ where $F\gg1$.

\emph{Spreading of the fluctuation front---}
To obtain a finite observable front, define $d_\eta(t)$ through a fixed fraction of the local stationary occupation:
\begin{equation}
{\begin{aligned}
\Pi(d_\eta(t),t) / \Pi_\infty=\eta, \qquad   
e^{-\frac{d_\eta(t)}{\ell_\mathrm{f}}}
\Big[1-e^{-F\big(d_\eta(t),t\big)}\Big]=\eta,
\end{aligned}}
\label{eq:front}
\end{equation}
with $0<\eta<1$.
We set $d_\eta(t)=0$ before $\Pi(0,t)$ reaches the chosen threshold. Since $F(d,t)\to\infty$ at each fixed $r$ when $t\to\infty$, the front approaches the finite value
\begin{equation}
{d_{\eta}(\infty)=\ell_\mathrm{f}\mathrm{log}(1/\eta).}
 \label{eq:dfinal}
\end{equation}
For $t\gg t_c$, the accumulated relaxation may be estimated as $F(d,t)\approx(t/\tau_0)e^{-d/\xi_\infty}$ (up to a finite early-time correction). At an unsaturated threshold position where $F(d_\eta,t)\ll1$, Eq.~\eqref{eq:front} implies
\begin{equation}
{d_\eta(t)\approx\ell_{\mathrm{eff}}
\mathrm{log}\!\left(\frac{t}{\eta\tau_0}\right),}
\label{eq:dlog}
\end{equation}
This result holds within the intermediate-time regime. As the threshold approaches the final exponential envelope, Eq.~\eqref{eq:front} describes the crossover to $d_{\eta}(\infty)$. Note that a relative-to-local-saturation front, defined by $F(d_c,t)=1$, instead behaves as $d_c(t)\approx\xi_\infty\mathrm{log}(t/\tau_0)$ and does not saturate on an infinite lattice.

\emph{RMSD growth---}
For a symmetric fluctuation distribution, the continuous approximation to the normalized root-mean-square distance is
\begin{equation}
{\sigma^2(t)=\frac{\displaystyle\int_0^\infty dr\,r^2\Pi(r,t)}{\displaystyle\int_0^\infty dr\,\Pi(r,t)}.}
\label{eq:sigma}
\end{equation}
To obtain a simple analytic intermediate-time estimate, we approximate the occupation distribution by $\Pi(d,t)\approx\Pi_\infty e^{-d/\ell_\mathrm{f}}\Theta[d_c(t)-d]$, where the crossover radius satisfies $F(d_c,t)\approx1$. The Integration gives
\begin{equation}
{\sigma^2(t)\approx\ell_\mathrm{f}^2\left[2-\frac{x^2+2x}{e^x-1}\right],\qquad x=\frac{d_c(t)}{\ell_\mathrm{f}}.}
 \label{eq:sigmacrossover}
\end{equation}
If the crossover is sufficiently sharp and $\xi_\infty\ll d_c(t)\ll\ell_\mathrm{f}$, the occupied region has an approximately uniform envelope and $\sigma(t)\approx d_c(t)/\sqrt{3}$, so
\begin{equation}
{\sigma(t)\approx\frac{\xi_\infty}{\sqrt{3}}\mathrm{log}\left(\frac{t}{\tau_0}\right).}
 \label{eq:sigmalog}
\end{equation}
This logarithmic scaling exists in an intermediate time window and it is not guaranteed for arbitrary $\xi_\infty/\ell_\mathrm{f}$. At long times, Eq.~\eqref{eq:Pi} approaches the final exponential distribution, and Eq.~\eqref{eq:sigma} yields
\begin{equation}
{\sigma_\infty=\sqrt{2}\,\ell_\mathrm{f}.}
 \label{eq:sigmafinal}
\end{equation}
Therefore, in this effective model, an initially vanishing development length naturally generates an effective onset delay, and the exponentially distance-dependent development rate permits intermediate logarithmic spreading.

\section{II. Supplementary numerical results}

\emph{Sampling number effect}---We study the effect of the sampling number of disorder on the RMSD dynamics. In Fig.~\ref{sampling_number}, we show the RMSD dynamics for different sampling numbers $N_{\mathrm{s}}$ and disorder strengths. For small $N_{\mathrm{s}}=10$, there are significant fluctuations in RMSD dynamics, particularly at large disorder strength. As $N_{\mathrm{s}}$ increases up to $100$ and $1000$, the result converges to a smooth curve. Thus, our choice $N_{\mathrm{s}}\geq100$ in the main text is reliable for capturing the RMSD growth.

\emph{Disorder-free case}---When disorder is absent, the quantum fluctuation is expected to grow quickly, since it manifests the error around mean-field dynamics. In Fig.~\ref{no_disorder}, we show the dynamics of $\overline{C(j,t)}$ at disorder-free case for $N=2$ and $L=500$. The growth of quantum fluctuation exhibits a perfect ballistic cone over time and space. Meanwhile, the RMSD grows linearly as $\propto t$ at late times, as shown in Fig.~\ref{no_disorder}(b). This can also be found in the profiles at different times, where two typical peaks spread out at a steady speed (Fig.~\ref{no_disorder}(c)). As shown in the main text, due to the presence of disorder, the spreading of quantum fluctuation is restricted and gradually turns into logarithmic growth. This manifests the suppression effect of disorder.

\emph{Profiles of quantum fluctuation}---In Fig.~\ref{profile_evolution}, we show the profiles of $\overline{C(j,t)}$ at different times and disorder strengths. When disorder is weak, the profile expands a little as time increases. But the profile always exhibits an exponential decay over space, corresponding to the logarithmic growth of RMSD. As the disorder is strong enough, the profiles stop spreading and almost coincide with each other, as shown in Fig.~\ref{profile_evolution}(d). This is in stark contrast to the disorder-free case.

\emph{Evolution of particle density}---In Fig.~\ref{density_evolution}, we show the evolution of particle density in the BH model and GP model, respectively. In the presence of strong disorder, starting from a condensed state, the particle density is expected to freeze at late times. As a mean-field description, the GP result should closely follow the BH result. As shown in Fig.~\ref{density_evolution}(a-f), despite different disorder settings, the particle density in the BH model spreads slowly and finally freezes over time. Similar behavior is found in the GP model (Fig.~\ref{density_evolution}(g-l)). As expected, under the same parameters, the particle density evolution in the GP model highly mimics that in the BH model. The quantum fluctuation in the main text appears from the difference in particle density in the two models.

\emph{Initial state with finite width}---To further support our findings, we consider another initial state with finite width as $\psi(t=0)=\frac{1}{\sqrt{3}}(\delta_{j_0-1}+\delta_{j_0}+\delta_{j_0+1})$. The dynamics of the corresponding quantum fluctuation for different disorder strengths are shown in Fig.~\ref{initial_state2_transport}. Overall, $\overline{C(j,t)}$ exhibits a similar spreading behavior as that in the main text. The spreading cone is still well bounded by the corresponding ballistic bound. What differs is that, due to the finite width of the initial state, the profile of $\overline{C(j,t)}$ is wider. We also show the RMSD dynamics for different disorder strengths in Fig.~\ref{initial_state2_sigma}. At early times, RMSD shows a sudden decrease due to the initial finite width. In both random and quasiperiodic settings, a logarithmic growth is present, and the slope drops algebraically as disorder strength increases. The quasiperiodic disorder exhibits a stronger suppression on quantum fluctuation than random disorder. 

\emph{Two-dimensional case}---We consider the 2D case of the same setup to test the ballistic bound and confirm the generality of logarithmic growth. Due to the exponentially large Hilbert dimension, we only consider $2$ bosons in a lattice of size $25\times25$. The results are shown in Fig.~\ref{2D_case}. One can find very similar dynamical behavior as in the 1D case. The ballistic bound at early times is still valid in the two disorder cases. The logarithmic growth is remarkable at intermediate times up to $Jt=100$, and the slope decays algebraically with increasing $\lambda$.

\emph{Long-time dynamics}---In Fig.~\ref{long_time}, we show the long-time dynamics (up to $Jt=10^3$) of the RMSD at $N=2$ for different disorder strengths $\lambda$. We find that the logarithmic growth is a transient behavior up to intermediate times $Jt\sim10^2$. At late times, the RMSD tends to saturate, especially for large disorders.

\emph{Convergence of TEBD calculations}---To ensure the validity of the TEBD method, we show the RMSD dynamics and the final spatial profile of the quantum fluctuation $\overline{C(j,t)}$ at $N=10$ for different bond dimensions $\chi$ in Fig.~\ref{tebd}(a-b). Clearly, the RMSD dynamics for $\chi=1024$ and $2048$ match each other and exhibit logarithmic growth at intermediate times. The spatial profiles also perfectly coincide with each other and decay exponentially, indicating the convergence of TEBD results at $\chi=1024$. Therefore, we use $\chi=1024$ as typical parameters in our calculations. In Fig.~\ref{tebd}(c-d), we also compare the exact results with TEBD results at $N=4$. In both random and quasiperiodic settings, the TEBD results are consistent with the exact results and properly reproduce the logarithmic behavior.

\end{twocolumngrid}

\end{document}